\documentclass[fleqn,usenatbib]{mnras}

\usepackage{newtxtext,newtxmath}

\usepackage[T1]{fontenc}

\DeclareRobustCommand{\VAN}[3]{#2}
\let\VANthebibliography\thebibliography
\def\thebibliography{\DeclareRobustCommand{\VAN}[3]{##3}\VANthebibliography}

\usepackage{graphicx}	
\usepackage{amsmath}	
\usepackage{placeins}
\usepackage{xcolor}
\usepackage{orcidlink}

\title[MIGHTEE Continuum DR1 Host Associations]{MIGHTEE: {The} Host-Galaxy Associated Catalogue of the Radio Sources in MIGHTEE Continuum Data Release 1 }

\author[C. L. Hale et al.]{C. L. Hale$^{\orcidlink{0000-0001-6279-4772}}$,$^{1, 2}$\thanks{E-mail: Catherine.Hale@ed.ac.uk}
J. Hamlett$^{\orcidlink{0009-0002-9611-1970}}$,$^{2}$
I. H. Whittam$^{\orcidlink{0000-0003-2265-5983}}$,$^{2, 3}$
M. J. Jarvis$^{\orcidlink{0000-0001-7039-9078}}$,$^{2, 3}$ 
M. J. Hardcastle$^{\orcidlink{0000-0003-4223-1117}}$,$^{4}$
S. L. Jung$^{\orcidlink{0000-0001-5512-3735}}$,$^{2}$  \newauthor 
N. Stylianou$^{\orcidlink{0000-0002-6029-005X}}$,$^{2}$  
R. G. Varadaraj,$^{2}$  
M. Vaccari$^{\orcidlink{0000-0002-6748-0577}}$,$^{5, 3, 6}$ 
L. Barchiesi$^{\orcidlink{0000-0003-3419-538X}}$,$^{7, 5, 6}$
P. N. Best,$^{1}$  \newauthor 
K. K. L. Charlton$^{\orcidlink{0000-0002-2925-2047}}$,$^{6, 8, 7}$   
E. Malefahlo$^{\orcidlink{0000-0001-8945-2879}}$,$^{9}$  
E. Moravec$^{\orcidlink{0000-0001-9793-5416}}$,$^{10}$ 
J. P. Moss,$^{11}$ 
M. Rakototafika,$^{7, 12}$    \newauthor 
T. F. Rarivoarinoro,$^{7, 5, 12}$ 
F. Sinigaglia,$^{13, 14, 15, 16}$   
M. N. Tudorache,$^{17}$  
S. I. Loubser$^{\orcidlink{0000-0002-3937-7126}}$,$^{18, 19}$  
L. Heino$^{\orcidlink{0000-0002-1282-4272}}$,$^{5, 7, 20}$  \newauthor 
A. Saintonge,$^{21, 22}$  
A. Mazumder,$^{23}$  
N. Netshiavha${\orcidlink{0000-0003-0295-4281}}$,$^{7}$ 
D. J. B. Smith$^{\orcidlink{0000-0001-9708-253X}}$,$^{4}$   
L. Verdes-Montenegro,$^{24}$   \newauthor 
J. Delhaize$^{\orcidlink{0000-0002-6149-0846}}$,$^{7}$
H. Pan,$^{25, 26}$   
F. Pozzi,$^{8, 27}$ 
Z. Randriamanakoto,$^{12, 28}$  
L. Stockenstroom$^{\orcidlink{0000-0001-7520-6503}}$,$^{7}$   \newauthor 
A. A. V\u{a}r\u{a}\c{s}teanu,$^{2}$  
E. Vardoulaki$^{\orcidlink{0000-0002-4437-1773}}$,$^{29, 30}$   
W. L. Williams,$^{3§}$ 
R. A. A. Bowler,$^{23}$ 
L. Marchetti$^{\orcidlink{0000-0003-3948-7621}}$,$^{7, 6, 5}$ \newauthor 
A. Matthews$^{32}$  
\\ \\
Affiliations are listed at the end of the paper
}

\date{Accepted XXX. Received YYY; in original form ZZZ}

\pubyear{2015}

\begin{document}
\label{firstpage}
\pagerange{\pageref{firstpage}--\pageref{lastpage}}
\maketitle

\begin{abstract} 
{Radio continuum surveys provide samples of active galactic nuclei (AGN) and star forming galaxies (SFGs) to high redshifts, free of biases due to dust obscuration. However, radio detected sources require multi-wavelength counterparts to understand their intrinsic properties (e.g. redshift, stellar mass) and to study the evolution of star formation and AGN activity. In this work we present host galaxy counterparts for the MeerKAT International GHz Tiered Extragalactic Exploration (MIGHTEE) survey continuum Data Release 1 in regions with the best ancillary data (totalling 7.5 sq. deg). We combine statistical cross-matching and visual inspection to identify $K_s$-band selected host galaxies, and additionally combine multiple radio components into single physical objects, where needed. This results in a combined radio catalogue of $\sim$66 000 sources, with host counterparts and redshifts identified for $\sim$95 per cent of sources in the COSMOS field, $\sim$91 per cent in XMM-LSS and $\sim$90 per cent in CDFS-DEEP. This includes a significant fraction of sources with spectroscopic redshifts within the COSMOS field ($\sim$50 per cent), with $\sim$30 and $\sim$20 per cent in the XMM-LSS and CDFS-DEEP fields respectively. Using the cross-matched catalogue, we make an initial identification of radio-excess and star forming galaxies based on comparisons of the radio luminosities to host star formation rates. Using this split as a proxy for radio loud AGN or SFGs, we present expectations for the redshift distributions of these sources, finding broad agreement with those from deep radio luminosity functions and simulated catalogues.}
\end{abstract}




\section{Introduction}
\label{sec:intro}
We are in an era of radio astronomy where a combination of deep and wide area surveys from telescopes such as the Low Frequency Array \citep[LOFAR;][]{vanHaarlem2013}, MeerKAT \citep{Jonas2009, Jonas2016}, {the} Australian Square Kilometre Array Pathfinder \citep[ASKAP][]{Hotan2021}, {the upgraded Giant Metrewave Radio Telescope \citep[uGMRT][]{uGMRT}} and the Karl G. Jansky Very Large Array (VLA) are unveiling a wealth of previously undetected radio sources across the Universe. Numerous surveys with these telescopes such as those presented by {\cite{Jarvis2016, Smolcic2017, Tasse2021, Sabater2021,vandervlugt2021, deJong2024} and \cite{Lal2025}}, have focussed on deep {observations} over areas where there is an abundance of multi-wavelength data. This allows such surveys to trace faint extragalactic radio sources over large periods of cosmic time and thus probe the evolution of radio source populations {and the properties of their host galaxies}. These deeper surveys are complemented by larger area surveys from {LOFAR, the VLA and ASKAP \citep{Lacy2020,Hopkins2025, Shimwell2026}} which can help to trace both rare and variable radio sources {as well as more diverse environments and more local sources}. Importantly, extragalactic radio sources can present a unique perspective of the extragalactic sky. Unlike optical/IR emission (where stars, gas and dust dominate the emission), radio emission (at $\lesssim$GHz frequencies) {predominantly traces non-thermal} synchrotron emission \citep[see e.g.][]{Condon1992}. This results in observations {that detect} two key extragalactic populations: active galactic nuclei (AGN) and star forming galaxies \citep[{SFGs;} see e.g.][]{MauchSadler2007, Smolcic2017b, Whittam2022, Best2023}. 

In AGN, relativistic electrons spiral in powerful jets {and the extended structures that they generate}, producing radio synchrotron emission. For radio loud AGN, there are two apparent accretion modes which play a role in their fuelling \citep[{High/Low excitation radio galaxies;} see reviews in e.g.][]{Heckman2014, Hardcastle2020}. Whilst initially such modes were believed to accrete at different Eddington rates \citep[see e.g.][]{Hardcastle2007, Best2012, Mingo2014}, evidence {from} deep radio surveys {has} suggested a much broader overlap for intrinsically less luminous populations \citep[see e.g.][]{Whittam2018, Whittam2022}. However, recent work using spectroscopic data has again indicated a more pronounced split between the populations \citep[see][]{Arnaudova2025} than previous {spectroscopic} studies \citep{Whittam2018}. Therefore, questions still remain and it is necessary to observe radio sources where deep ancillary data {are} available to provide information about the properties of host galaxies. {At the same time, for} radio quiet AGN, the depth of recent surveys has allowed a significant increase in {the numbers which have been observed}, {helping to advance studies into} whether jets, star formation, winds or other AGN processes are dominating their emission \citep[see e.g.][]{Laor2008, White2015, Chen2023, White2025, Njeri2026, Jackson2026}. 

For SFGs, the synchrotron emission originates from high energy cosmic electrons spiralling in the magnetic fields of supernova remnants. As such, radio {emission from SFGs is} known to be an excellent (delayed) tracer of star formation \citep[see e.g.][]{Garn2009, Davies2017, Smith2021, Matthews2021b, Cochrane2023, Cook2024, Thykkathu2026}, crucially without the effects of dust obscuration which can plague other wavelengths \citep[see discussions in][]{Madau2014}. This can allow star formation to be traced across large periods of the Universe's history, provided the surveys are deep enough. Using the radio emission as a star formation rate (SFR) tracer across cosmic time is one of the science goals for future deep radio surveys with telescopes such as LOFAR2.0 and the Square Kilometre Array Observatory (SKAO). However, considerations may have to be made to account for the effects of inverse Compton scattering from the cosmic microwave background (CMB), which could affect the inference of star formation rates at high redshifts \citep[see e.g.][]{Murphy2009, Whittam2025}.

Observing both AGN and SFGs is essential to understand how and why galaxies evolve over cosmic time. This is because both star formation and black holes are known to play crucial roles in {galaxy properties and their evolution} \citep[see e.g.][for reviews]{Benson2010, Kormendy2013, Heckman2014, Hardcastle2020}, {and affecting} their growth through the effects of {feedback and fuelling}. Therefore, radio surveys are crucial to build a comprehensive picture of galaxy evolution. However, continuum radio surveys alone are unable to provide the critical source information such as the redshift ($z$) of {a source} and the stellar mass of its host galaxy ($M_{*}$) which are crucial to study how radio sources evolve \citep[see e.g.][]{Novak2017, Smolcic2017c, Smith2021, Kondapally2022, Cook2024, Kondapally2025}. { Determining} {the physical properties first} requires the identification of host galaxies {in order to user photometric and spectroscopic multi-wavelength data to measure properties of the radio sources through methods such as SED modelling} \citep[see e.g.][]{Smolcic2017b, Whittam2022, Best2023, Das2024, Drake2024, Arnaudova2025}. Whilst identifying host galaxies may sound like a simple task, there is typically a significant mis-match in the angular resolution of radio surveys to optical/near-IR surveys, {with radio generally having poorer resolution \citep[though see the high-resolution, but small area COSMOS-XS survey;][]{vandervlugt2021}}. This can lead to complexity when identifying host galaxies - especially when multiple optical galaxies may exist within the extent of the restoring beam of the radio data. Moreover, radio sources can be extended and composed of physically separated {areas of} radio emission. This can lead to traditional source finders \citep[e.g.][]{Selavy, PyBDSF} struggling to associate all the flux into a single physical source {(thus inaccurately measure intrinsic properties) and {may struggle to identify a} radio position which is close to the host galaxy. Therefore, the analysis of numerous deep radio surveys has {focused} on galaxy identification and classification to maximize the science capabilities of radio data \citep[e.g.][]{Smolcic2017b, Algera2020, Kondapally2021, Whittam2022, Best2023, Whittam2024}.} 

{One} deep radio survey which is transforming views of the radio sky is the MeerKAT International GHz Tiered Extragalactic Exploration (MIGHTEE) survey \citep{Jarvis2016}. As one of the large survey projects with MeerKAT, its aims span continuum, H{\sc i} and polarisation science with the over-arching goal of studying galaxy evolution over large periods of cosmic time. It specifically observes four extragalactic fields, each of which {has} an abundance of multi wavelength data across the electromagnetic spectrum. Initially 5 sq. deg spanning the {Cosmic Evolution Survey field \citep[COSMOS, 1.5 sq. deg][]{Scoville2007}} {and} the XMM-Newton Large-Scale Structure field (XMM-LSS, 3.5 sq. deg) were released in an early science data release \citep[MIGHTEE-ES;][]{Heywood2021}, with a subset of the COSMOS field ($\sim$0.8 sq. deg) cross-matched to host galaxies \citep{Whittam2024} and classified into AGN and SFGs through a combination of diagnostics \citep{Whittam2022}. {The} area of MIGHTEE was significantly increased to $\sim$20 sq. deg in Data Release 1 \citep[hereafter MIGHTEE-DR1;][]{Hale2025}. This release included areas spread across three of the four deep fields which MIGHTEE will observe. Two of the fields, COSMOS (4.2 sq. deg) and XMM-LSS (14.4 sq. deg) represent the final L-band images and the third field was a singular deep pointing over the Chandra Deep Field South field (denoted here as CDFS-DEEP\footnote{{Throughout the paper we use CDFS-DEEP to refer to the MIGHTEE-DR1 coverage in the field and CDFS to refer to the larger field area with multi-wavelength coverage.}}, 1.5 sq. deg). The {L-band} radio coverage will be extended in the future with larger coverage over CDFS and imaging of the fourth field, the European Large Area ISO Survey S1 (ELAIS-S1) field. 

In this work we identify the host galaxies and redshifts for radio {continuum} detected sources in MIGHTEE-DR1. The paper is arranged as follows. First, in Section \ref{sec:data} we introduce the data used in this work, both for the initial radio source catalogues and the multi-wavelength catalogues that we cross-match to. In Section \ref{sec:methods} we introduce the methods used to cross-match our radio sources to such host galaxies through both statistical and visual-cross matching. We present our cross-matched catalogue and associated analysis in Section \ref{sec:results} before discussing the results of initial investigations into the host properties of the radio sources in Section \ref{sec:hostproperties}. Finally, we summarise and draw conclusions in Section \ref{sec:conclusions}. Unless otherwise stated, we assume a concordance cosmology of $H_0=70$~km~s$^{-1}$~Mpc$^{-1}$, $\Omega_{\textrm{M}}=0.3$ and $\Omega_{\Lambda}=0.7$ and we assume a radio spectral index, $\alpha=0.7$ where the flux density, $S_{\nu}$, is related to frequency, $\nu$ as $S_{\nu} \propto \nu^{-\alpha}$. {Magnitudes quoted are in the AB system \citep{Oke1983} and a Chabrier initial mass function \citep[IMF;][]{Chabrier2003} is adopted.}

\section{Data}
\label{sec:data}

\subsection{Radio Data}
\label{sec:mightee_data}
The base radio catalogue for this work comes from the L-band ($\sim$1.3 GHz) MIGHTEE-DR1 radio catalogue from \cite{Hale2025}. Each observation was reduced using the \texttt{oxkat} pipeline \citep{Oxkat}, using a combination of steps to calibrate the data, accounting for both direction independent and direction dependent calibration. Imaging was performed using two Briggs' weighting values \citep{Briggs1995}. These Briggs' weightings produced both a higher (5 {arcsec}) resolution image, which {is} typically less sensitive, as well as a lower (7-9 {arcsec}) resolution image, which {is} typically more sensitive. Whilst the CDFS-DEEP field observations were all centred on a single pointing, multiple pointings were mosaiced together for the XMM-LSS and COSMOS fields to produce images of larger contiguous regions. Each image {has} rms levels of $\sim1-5$ $\muup$Jy/beam, where a combination of both thermal noise and confusion noise \citep[see e.g.][]{Condon1974} contributed to the background rms level, with source confusion becoming more significant for the lower resolution images. {For CDFS-DEEP the median rms is $\sim 2 \muup$Jy/beam in both images, this increases in the COSMOS and XMM-LSS fields to $\sim 3 \muup$Jy/beam in the lower resolution images and $\sim 5 \muup$Jy/beam in the higher resolution images. The full details of this, alongside the improved rms in the central regions are detailed in Table 1 of \cite{Hale2025}.} Alongside each continuum image, an associated effective frequency map was also released to reflect the variation in frequency observed across the field of view due to a combination of effects such as differences in the primary beam with frequency and differences in flagging per individual observation. 

Source catalogues were extracted using the Python Blob Detection Source Finder \citep[\texttt{PyBDSF};][]{PyBDSF}. {A} multi-stage approach to run \texttt{PyBDSF} {was used} to identify an initial list of sources and a residual map, with sources removed. {The residual map} was used to generate an rms map which was then supplied back to \texttt{PyBDSF} to detect sources above a peak flux detection limit of 5$\sigma$, and modelling the emission as Gaussian components. Further details of the source finding process {are given} in \cite{Hale2025} and the total number of sources (and Gaussian components) are presented in Table 1 of \cite{Hale2025}. {\cite{Hale2025} performed no source verification}, therefore whilst the majority of sources are genuine, single sources, it was unclear what was the contribution of (i) physical sources split into multiple \texttt{PyBDSF} sources, or (ii) blended sources in the catalogue. {The contribution of such sources are expected to be relatively small - the most comparable data which has similar depth, resolution and uses the same source finder is the LoTSS Deep Fields. In their work, \cite{Kondapally2021} find $\sim$1-2 per cent of \texttt{PyBDSF} sources were merged into a single sources. However,} the motivation of the current paper is to produce a catalogue with radio components associated to a single object, and obtain host galaxy positions and redshifts for the combined radio catalogue, where available.

{For the radio catalogue we use the high resolution ($\sim$5 {arcsec}) \texttt{PyBDSF} source catalogues of \cite{Hale2025}. This is to reduce the issues arising from source confusion, the blending of radio sources, and the chances of having multiple multi-wavelength sources within the radio extent. }

\subsection{Multi-wavelength Catalogues}
\label{sec:optIR_data}
As discussed in Section \ref{sec:intro}, the MIGHTEE fields were selected to cover four extragalactic fields which have an abundance of multi-wavelength data. The coverage and depth of this multi-wavelength data varies per field, but includes data from X-ray \citep[e.g.][]{Xcosmos, XMMServs, Xcdfs}, UV \citep[e.g.][]{Martin2005, Morrissey2007}, optical \citep[e.g.][]{Aihara2022, Vaccari2016} {and} IR \citep[e.g.][]{Sanders2007, Mauduit2012, Jarvis2013, McCracken2012, Lacy2021} wavelengths, and see compilations {of this multi-wavelength data} in e.g. \cite{HELP}. This is alongside additional radio observations {available} from surveys such as {those of} \cite{Tasse2008, Schinnerer2010, Hales2014, Smolcic2017, Hale2019, Heywood2020} and \cite{Lal2025}. For the association of host galaxies in this work we focus on $K_s$ band selected catalogues, {as this band} is often used for measuring stellar mass in galaxies \citep[see e.g.][]{Kauffmann1998} {and deep imaging covered large areas of the radio imaging}. For this work, we adopt {the} $K_s$ data from UltraVISTA \citep{McCracken2012}\footnote{We note {that} at the time of starting the cross-matching, this used catalogues which were derived from Data Release 5 (DR5) of UltraVISTA. The COSMOS UltraVISTA based catalogues have now been updated to include the final data release (DR6). {The catalogues which we use to obtain photometric redshifts (see Section \ref{sec:redshifts}) use DR6 for COSMOS and for CDFS/XMM-LSS updated catalogues were made to allow for better mosaicing. This does not change the majority of our cross-matching, but did lead to a need to match to the updated catalogues when we associated photometric redshifts. This will be discussed in Section \ref{sec:redshifts}.}}  in the COSMOS field and from the VISTA Deep Extragalactic Observations \citep[VIDEO;][]{Jarvis2013} survey in the XMM-LSS and CDFS fields. Details of the near-IR catalogues can be found in {\cite{Varadaraj2023}}. To summarise, \texttt{SExtractor} \citep{Bertin1996} was used to extract catalogues from the $K_s$ band, recording both aperture and auto magnitudes. The $K_s$ band catalogue was then used as a reference catalogue {and used \texttt{SExtractor} in dual-image mode to obtain forced photometry across other optical and IR wavelengths.}

In order to use these catalogues for statistical matching, we first {needed to ensure that} these catalogues are robust and have uniform detection across the field of view. Therefore, 5$\sigma$ magnitude limits were applied to the catalogues, using a limiting 5$\sigma$ $K_s$-band magnitude in 2 {arcsec} apertures of 24.8, 23.8 and 23.7 in the COSMOS, XMM-LSS and CDFS fields respectively. Additionally, as we require a clean sample of galaxies, we {removed} sources which are likely to be stars from the $K_s$ band catalogues before matching to the radio catalogues. We {adopted} the stellar locus method of {e.g. \cite{Jarvis2013, Hatfield2022}} to identify and remove potential stars using colour cuts on the \textit{g}-\textit{i} and \textit{J}-$K_s$ colours, using UltraVISTA, VIDEO and  Hyper Suprime-Cam Subaru Strategic Program \citep[HSC-SSP;][]{Aihara2022} {in 2 arcsec apertures.}

\subsection{Additional Masking across the Field of View}
\label{sec:masking}
\begin{figure}
    \centering
    \includegraphics[width=0.85\linewidth]{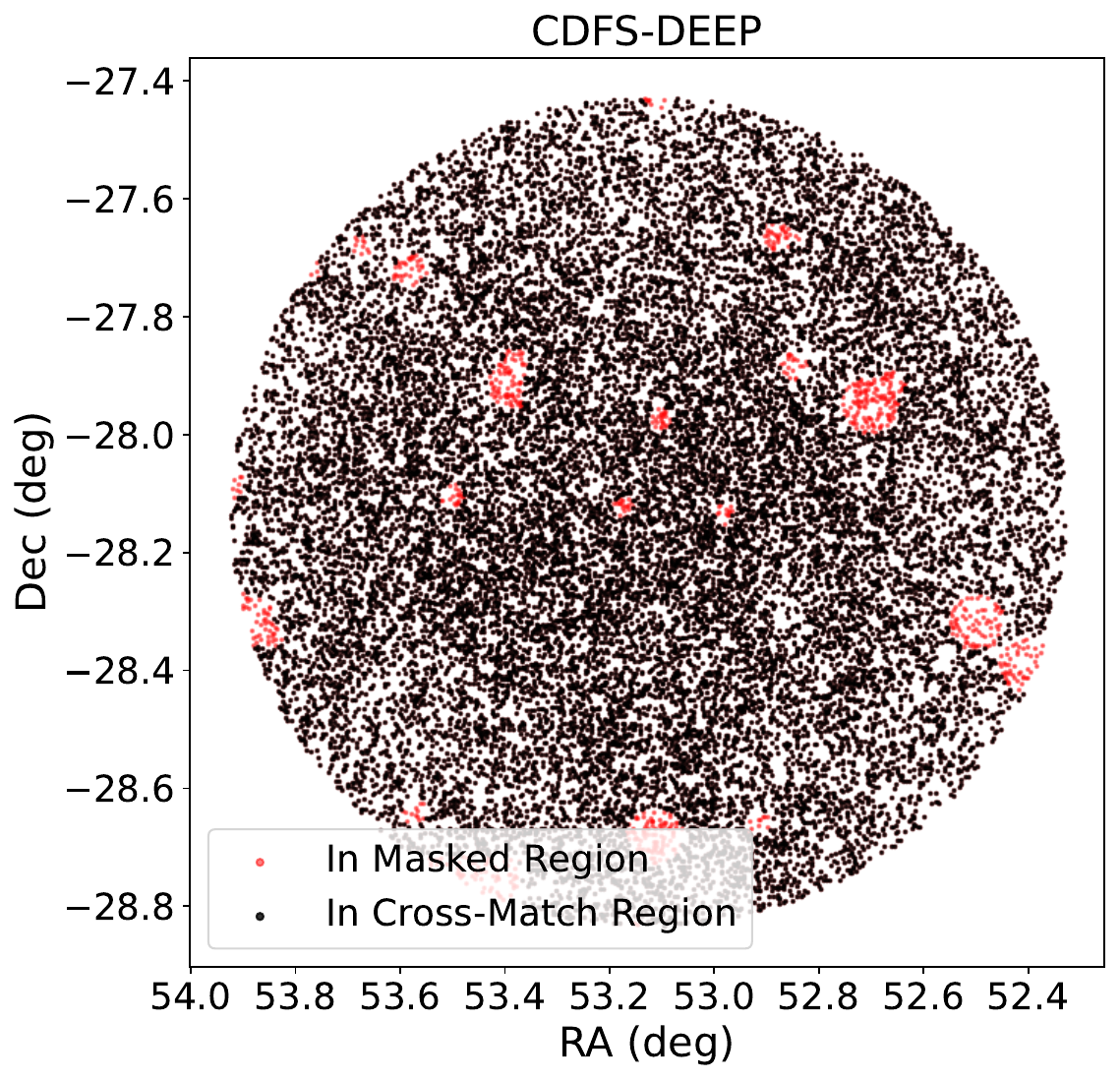} \\
    \includegraphics[width=0.85\linewidth]{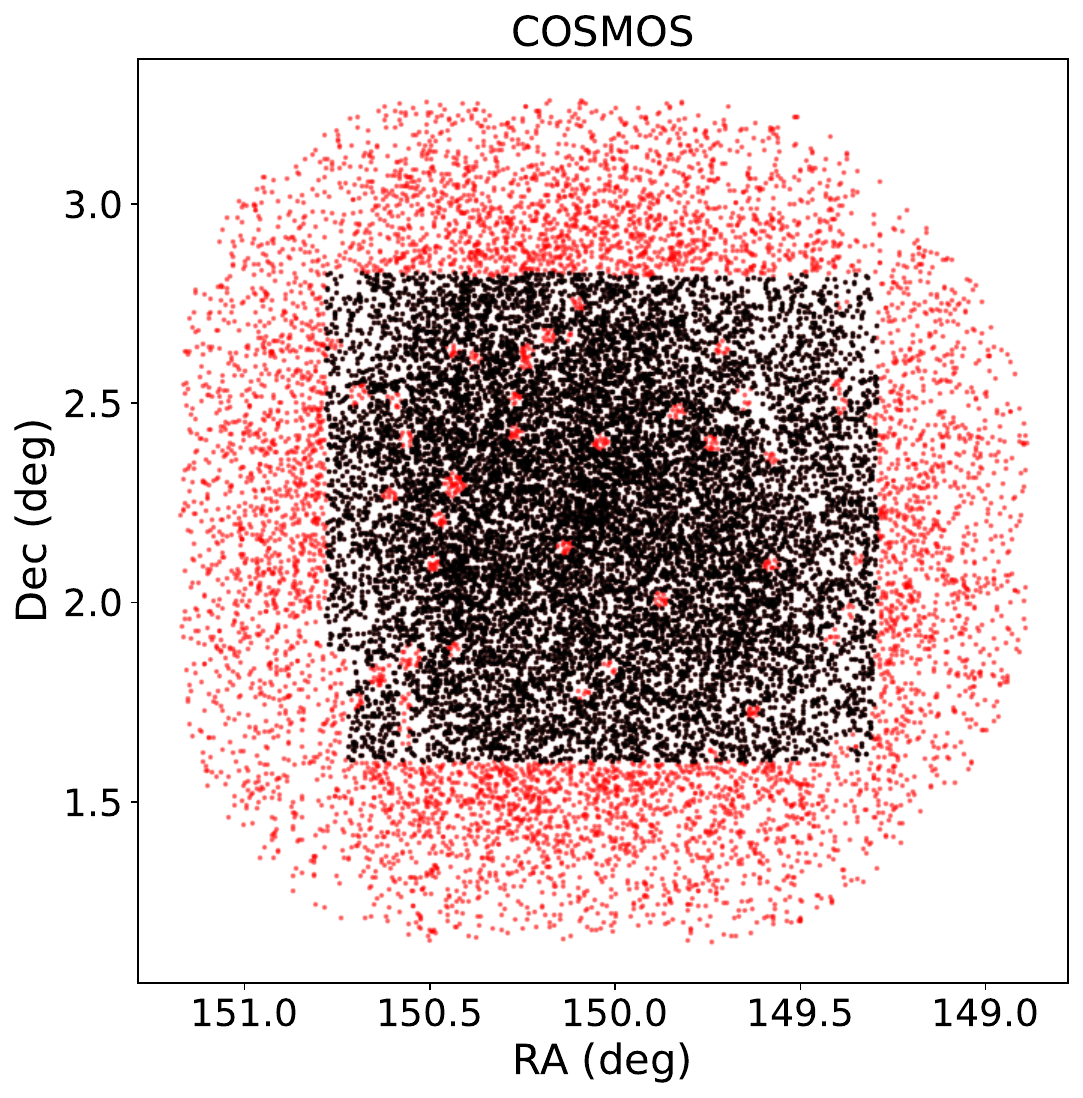} \\
    \includegraphics[width=0.85\linewidth]{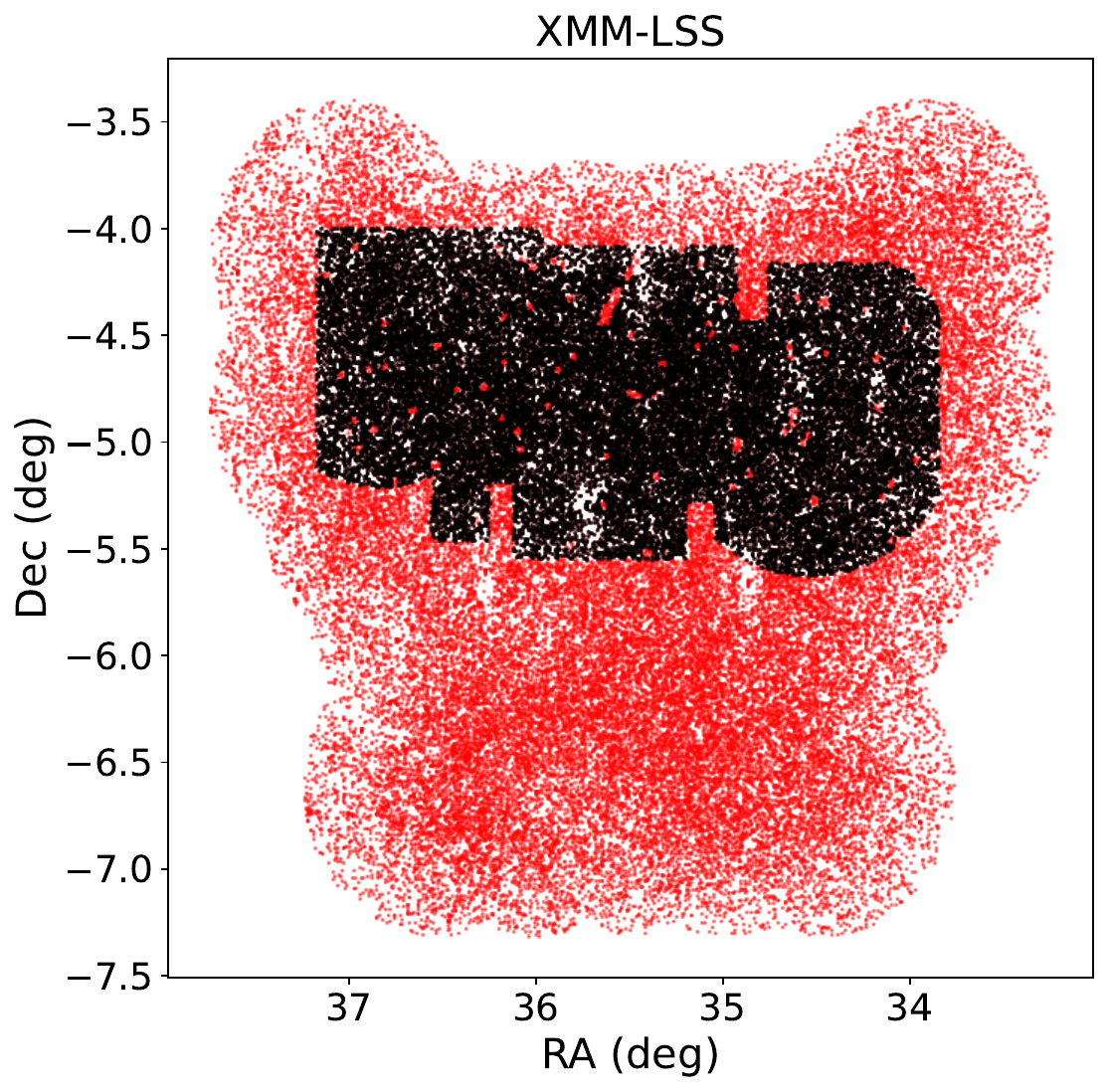} \\
    \caption{Full MIGHTEE-DR1 source catalogue from \protect \cite{Hale2025} with those radio sources in masked regions (and thus not cross-matched) shown in red, and those within the multi-wavelength regions used for cross-matching shown in black.}
    \label{fig:masking}
\end{figure}

Finally, we restrict {cross-matching our radio sources to the regions with the best multi-wavelength data and remove regions where image quality is affected by bright stars {(e.g. stellar ghosting) and other artefacts in the fields, as have been implemented in the $K_s$ band catalogues}. We applied spatial masks whose effect can be seen in Figure \ref{fig:masking}. The majority of the MIGHTEE CDFS-DEEP data is unmasked. For the COSMOS and XMM-LSS fields, instead, the MIGHTEE data covers an extended region significantly beyond where the best multi-wavelength data exists. Therefore, the area over which we can cross-match radio data to $K_s$ band hosts is reduced. For COSMOS, this is driven by the coverage of the UltraVISTA data and for XMM-LSS and CDFS it is driven by the combination of the VIDEO and HSC-SSP data. Combined, this results in a reduction of sources as shown in Table \ref{tab:maskingdata} {and results in an area with both the multi-wavelength and radio coverage totalling $\sim$1.5 sq. deg in CDFS-DEEP, $\sim$1.7 sq. deg in COSMOS and $\sim$4.3 sq. deg in XMM-LSS.}} \\

\noindent After applying the masking and reducing contamination from stars, we proceed to cross-match to the radio sources to identify their host galaxies. 

\section{Cross-Matching Techniques in Radio Surveys}
\label{sec:methods}
{Traditionally several methods {have been adopted}} to identify the host galaxies of radio sources, with pros and cons to each approach. The quickest and simplest is to adopt a simple positional match between the radio and {potential host galaxies}, using a positional match radius which is typically comparable to (or slightly smaller) than the beam size of the radio data. This works well with compact and isolated sources, and when the radio {resolution and that of the multi-wavelength imaging being matched to} are comparable. However, when the resolution of radio data is poorer or where the multi-wavelength data to be cross-matched to has a much higher source density, statistical methods are more appropriate for cross-matching compact objects. One such statistical method that is commonly used is the likelihood ratio \citep[LR;][see Section \ref{sec:lr}]{Sutherland1992}. This uses the magnitude distribution of galaxies (and those hosting radio sources) combined with positional information to identify probabilistic host matches, and {this technique has} been adopted for numerous radio catalogues \citep[see e.g.][]{McAlpine2012, Vernstrom2016, Kondapally2021}. 

However, probabilistic methods typically only work successfully for compact objects, but can encounter challenges for resolved sources, such as large, jetted AGN which may have been measured as separate objects by the radio source finder. Therefore, to {identify} a host galaxy, all the components of the source need to be associated with a single physical object. For these sources visual inspection is often used to identify the host galaxy {\citep[see e.g.][]{Banfield2015, Williams2019, Kondapally2021, Wong2025}}. However, visual inspection is a time consuming process, and so filtering the sources which should be visually inspected is useful \citep[see e.g.][]{Kondapally2021, Whittam2024}. 

For the MIGHTEE-ES data, each Gaussian component {in the lower resolution image} was visually inspected for a subset of the COSMOS field \citep[covering $\sim$0.8 sq. deg][]{Whittam2024}. This resulted in a catalogue of cross-matched sources for $\sim$80 per cent of sources in that region. Additionally, the host galaxies of the sources were also associated using the likelihood ratio technique, where $\sim$94 per cent of radio sources were identified as having a host galaxy using the likelihood ratio technique. They also used this to assess the agreement between the visually inspected and likelihood ratio associated host galaxies. For those sources determined to have a `good' match {(i.e. LR above the relevant threshold and reaching their agreement criteria from visual inspection)} in both the visually inspected and likelihood ratio catalogue, {there was 95.5 per cent agreement between the host galaxies}. This suggests that the likelihood ratio can be used robustly in a number of cases, but there will be sources for which visual inspection is required (see Section \ref{sec:lr_or_zoo}).

\subsection{Likelihood Ratio}
\label{sec:lr}
The {likelihood ratio (LR)} describes the probability that {two sources are related (in this case using the radio sources and the $K_s$ selected catalogue as potential hosts) compared to them being unrelated}. This probability takes into account both the proximity of potential host sources to the radio source as well as host galaxy properties. {One common method is to consider the magnitude distribution of the potential host galaxies and the background population, which is what we adopt in this work}. In this case, the likelihood ratio is given by:

\begin{equation}
\textrm{LR} = \frac{q(m) f(r)}{n(m)}.
\end{equation}

\noindent Here, $f(r)$ is the radial distribution of $K_s$ sources around radio sources, thus taking into account host proximity. The remaining factor ($q(m)/n(m)$) takes the ratio of the magnitude ($m$) distribution for $K_s$ sources which are true counterparts of the radio sources ($q(m)$) and normalises this by the magnitude distribution of the complete $K_s$ source catalogue used ($n(m)$). {To find $q(m)$ a maximum radius around the radio sources is adopted which accounts for the beam size as well as positional errors and calibration uncertainties. The details of this are available in the source code\footnote{{from \url{https://github.com/lmorabit/likelihood_ratio}}}.} 
{By} assigning a numerical value for the LR, it is possible to select the best match for the radio source and identify a potential host $K_s$ band galaxy. 

We obtain the likelihood ratio of potential host galaxies following the method of {works such as \cite{McAlpine2012} and} \cite{Whittam2024}, using an existing likelihood ratio matching code. The input $K_s$ and radio catalogues have the magnitude and stellar cuts and spatial masking applied, as {discussed in Section \ref{sec:masking}}. The best $K_s$ band match is considered suitable if the LR is above a threshold. As adopted for the MIGHTEE-ES data \citep{Whittam2024}, this threshold is calculated using the method of \cite{Williams2019} by balancing the completeness and reliability of matches. The {completeness, $C(L_{\textrm{thr}})$, and reliability, $R(L_{\textrm{thr}})$,} at a given likelihood ratio are defined as:

\begin{equation}
    C(L_{\textrm{thr}}) = 1 - \frac{1}{Q_0 N_{\textrm{radio}}}  \sum_{\textrm{LR}_i<L_{\textrm{thr}}} \frac{Q_0 \ \textrm{LR}_i}{Q_0 \ \textrm{LR}_i + (1-Q_0)},
    \label{eq:comp}
\end{equation}

\noindent and
 
\begin{equation}
    R(L_{\textrm{thr}}) = 1 - \frac{1}{Q_0 N_{\textrm{radio}}}  \sum_{\textrm{LR}_i<L_{\textrm{thr}}} \frac{1 - Q_0}{Q_0 \ \textrm{LR}_i + (1-Q_0)}.
    \label{eq:reliability}
\end{equation}

\noindent {The LR threshold for determining if a host is a suitable match is taken as the value of LR where the completeness and reliability curves intersect\footnote{{The threshold LR for the source catalogue are 0.25, 0.14 and 0.32 for the CDFS-DEEP, COSMOS and XMM-LSS fields respectively, increasing to 0.35, 0.35, and 0.47 for the Gaussian catalogues.}}.} In Equations \ref{eq:comp} and \ref{eq:reliability} {$Q_0$} is a normalisation fraction used to indicate the number of radio sources {expected to have} a match compared to the total number of radio sources ($N_{\textrm{radio}}$). {It is calculated in the code adopting the estimation method of \cite{Fleuren2012} and we impose a 5 arcsec maximum radius to fit this. For the three fields, we find $Q_0$ values for the source catalogues of $\sim$0.91 in CDFS-DEEP, $\sim$0.96 in COSMOS and $\sim$0.92 for XMM-LSS - indicating that we expect to have between 91-96 {per cent} of sources in the three fields to have associated host galaxies (and is similar to the results found, see Table \ref{tab:Nmatches}). {In general, for sources with a LR above the given threshold we adopt the host galaxy} (but see Section \ref{sec:lr_or_zoo}), thus reducing the need for a significant number of sources to be visually inspected.  }

\begin{figure*}
    \centering
    \includegraphics[width=\textwidth]{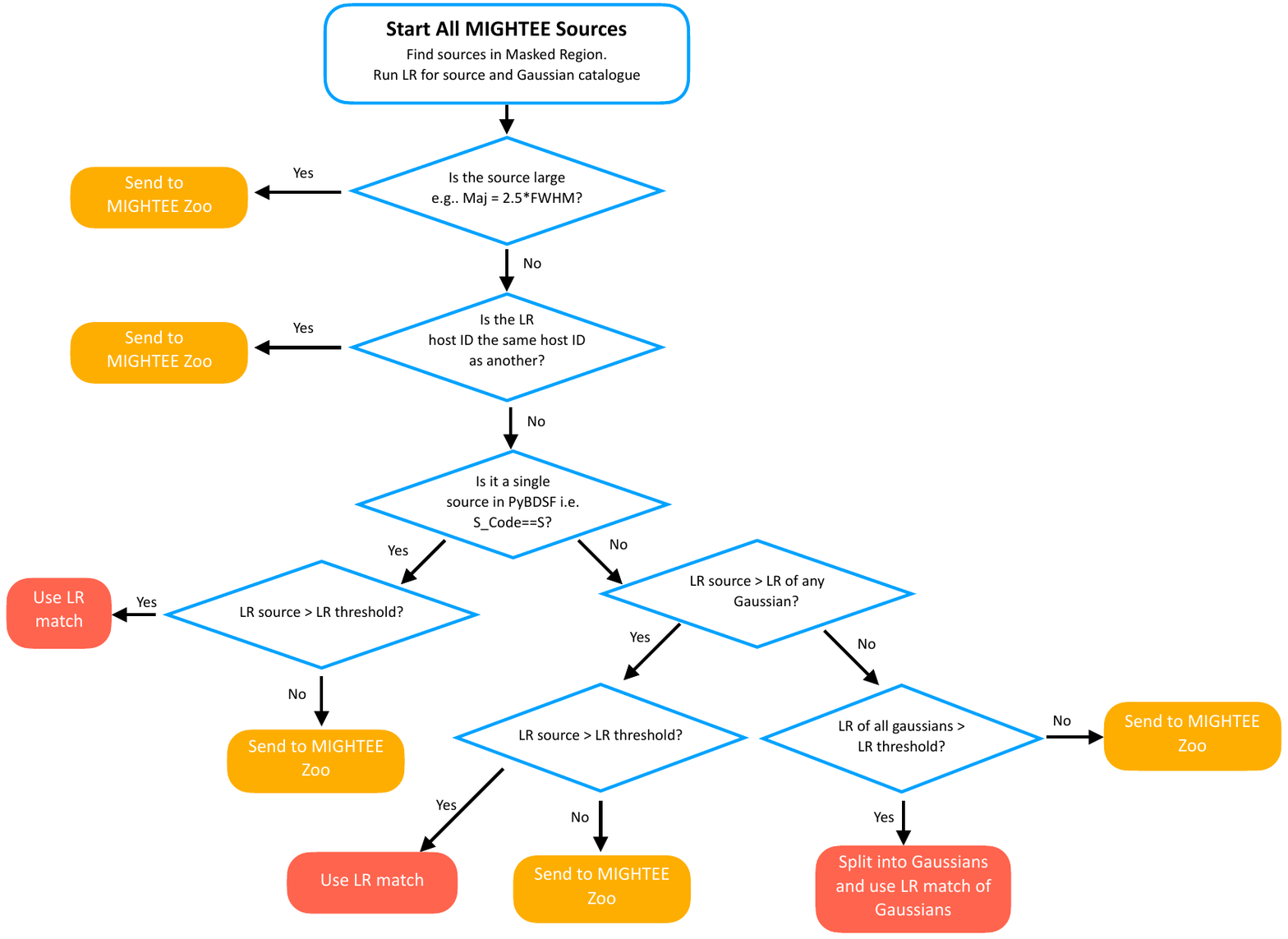}
    \caption{Decision tree flow chart used to determine whether the likelihood ratio matched source is suitable to associate a host galaxy from the near-IR or if the source should be sent to MIGHTEE zoo for {visual inspection}, see Section \ref{sec:lr_or_zoo}.}
    \label{fig:flowchart}
\end{figure*}

\subsection{Likelihood Ratio or Visual inspection?}
\label{sec:lr_or_zoo}
As we scale to the $\sim$65 000 sources in the {high-resolution} MIGHTEE-DR1 data {within the multi-wavelength regions} (as shown in Figure \ref{fig:masking}), visual inspection of each source (as done in MIGHTEE-ES) becomes impractical. As the early science data suggested {that} the LR can be trusted in a large number of cases, we use a process  of adopting the likelihood ratio match for a source where possible, and using further visual inspection when we believe {that} the LR may not be as robust. We construct a decision tree \citep[similar to {that of}][]{Kondapally2021} to take a series of steps to determine whether to adopt the LR match or send a source for visual inspection. {This is presented in Figure \ref{fig:flowchart}, and to summarise the steps are: } \\

\noindent (i) First, sources which are defined as large (having a non-deconvolved major axis, Maj$\geq 2.5 \times$ the {radio image's FWHM}) are automatically sent for visual identification. These sources might constitute lobed AGN or nearby star forming galaxies and so are likely to need combining with other radio objects. \\
    
\noindent (ii) Secondly, we {select} sources which have a LR match host ID which is the same as that of another {radio} source; these are automatically sent for visual inspection. \\
    
\noindent (iii) Next, we take two approaches based on whether a source is identified as a single Gaussian component source by \texttt{PyBDSF} (S\_code=`S') or not. For sources identified as a single component source and where the LR is above the threshold then this host galaxy is adopted. If not, the radio source is instead sent for visual inspection. \\

\noindent (iv) For those sources which have multiple Gaussian components we consider the LR of all Gaussian components within the source ({using a second LR catalogue which used the Gaussian catalogue instead of the radio source catalogue as input}). If the LR of the source is greater than the LR of any of the individual Gaussian {components and the source LR} is greater than the associated LR threshold, then the LR match for the overall source is adopted and the source is not sent for visual inspection. However, if the LR of the source is below the threshold, the source is sent for visual inspection. \\
    
\noindent (v) Finally, for those sources which have multiple Gaussian components where the LR of all Gaussian components are greater than the LR of the source then that might indicate a blended source. Therefore, if the LR of all the Gaussian components are greater than the respective threshold then the sources are de-blended into their Gaussian components and the LR host is adopted for each of these. Else, the source is sent for visual inspection. \\

\noindent Combined, these steps allow us to identify which sources should be sent for visual inspection through the `MIGHTEE Zoo' (see Section \ref{sec:zoo}). The number of sources which are sent to {MIGHTEE Zoo} for each field\footnote{{We note an update to the LR code which was performed after the COSMOS MIGHTEE zoo was started and prior to the other fields. In the table we include the numbers of source from the flowchart of the updated code, which was predominately adopted to classify sources, but include in brackets the number of sources which were sent to the zoo in reality. This was more sources than necessary from the flow chart, but as these were visually inspected, it should not affect the results.}} in each step outlined above (and the total numbers across all steps) are given in Table \ref{tab:maskingdata}.

\begin{table*}
  \centering
  \caption{Details of the number of radio sources used throughout the analysis after applying different steps of the cross-matching process. Included are the number of sources ($N_s$) or {Gaussians} ($N_g$) across the full MIGHTEE-DR1 images and over the restricted masked regions indicated in Figure \ref{fig:masking} (in red). {Also included are details of the number of sources which should be sent to MIGHTEE zoo or for which the likelihood ratio cross-matching could be used, using the different criteria from the flowchart in Figure \protect \ref{fig:flowchart}. For the COSMOS field, see the note at the end of Section \ref{sec:lr_or_zoo}.}  }
\begin{tabular}{l|rrrrrr}
\hline
& CDFS-DEEP & COSMOS & XMM-LSS & Total \\ \hline \hline 
N$_s$ Full Radio Image & 21 152 & 20 886 & 72 187 & 114 225 \\ 
N$_g$ Full Radio Image & 22 660 & 22 420 & 76 567 & 121 647 \\ 
N$_s$ Multi-wavelength region & 20 335 & 13 781 & 31 491 & 65 607 \\ 
N$_g$ Multi-wavelength region & 21 757 & 14 581 & 33 247 & 69 585 \\ \hline 
Large N to zoo & 1 870 & 1 380 & 2 891 & 6 141 \\ 
Sources with same Multi-wavelength ID & 135 & {15} & 198 & 353 \\ 
Single Sources to zoo & 1 057 & {238} & 1 530 & 3 120 \\ 
Single Sources use LR & 13 360 & {10 395} & 23 490 & 46 942 \\ 
Multi-component Sources use LR & 3 301 & {1 508} & 2 899 & 6 623 \\ 
Multi-component Sources use Gaussians LR & 278 & {148} & 164 & 1 585 \\ 
Multi-component Sources {to zoo} & 334 & {102} & 319 & 843 \\ \hline 
Total sources to use LR & 16 661 & {11 902} & 26 389 & 53 565 \\ 
Total sources to use Gaussians & 278 & {148} & 164 & 1 585 \\ 
Total sources {to MIGHTEE} Zoo & 3 396 &  {1731 (2 123)} & 4 938 & 10 457 \\ \hline 
\end{tabular}
\label{tab:maskingdata}
\end{table*}

\begin{figure*}
    \begin{minipage}[t]{.24\textwidth}
        \centering
        \includegraphics[width=0.95\textwidth]{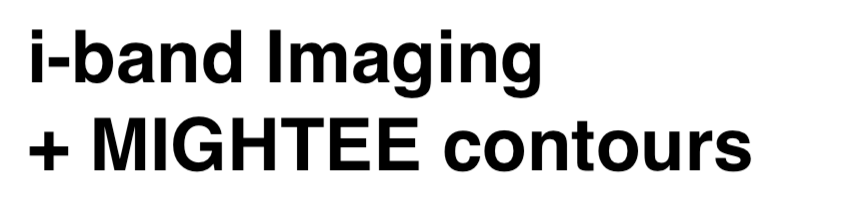}
    \end{minipage}
    \begin{minipage}[t]{.24\textwidth}
        \centering
        \includegraphics[width=0.95\textwidth]{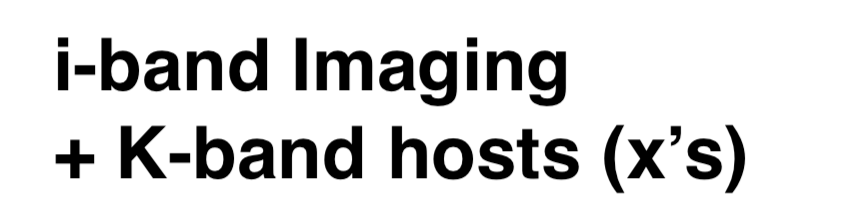}
    \end{minipage}  
    \begin{minipage}[t]{.24\textwidth}
        \centering
        \includegraphics[width=0.95\textwidth]{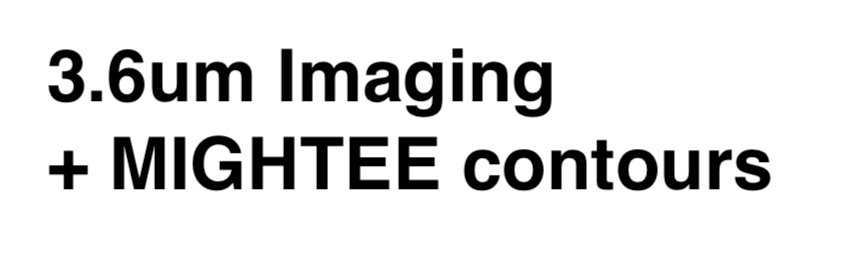}
    \end{minipage}
    \begin{minipage}[t]{.24\textwidth}
        \centering
        \includegraphics[width=0.95\textwidth]{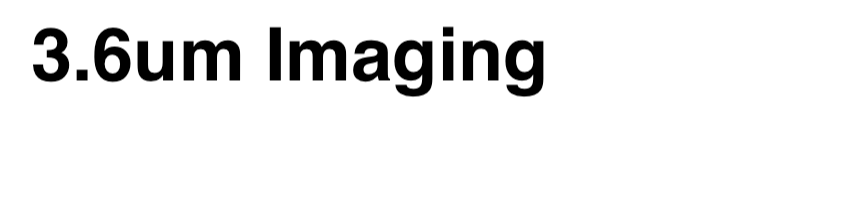}
    \end{minipage}  
    \linebreak
    \begin{minipage}[t]{.24\textwidth}
        \centering
        \includegraphics[width=0.95\textwidth]{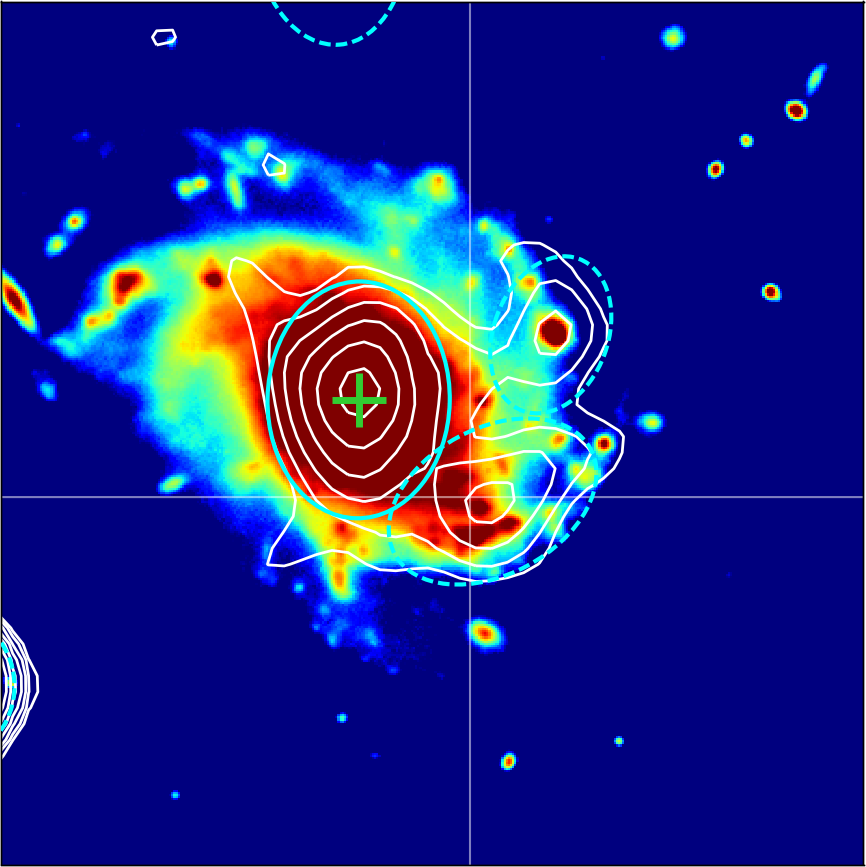}
    \end{minipage}
    \begin{minipage}[t]{.24\textwidth}
        \centering
        \includegraphics[width=0.95\textwidth]{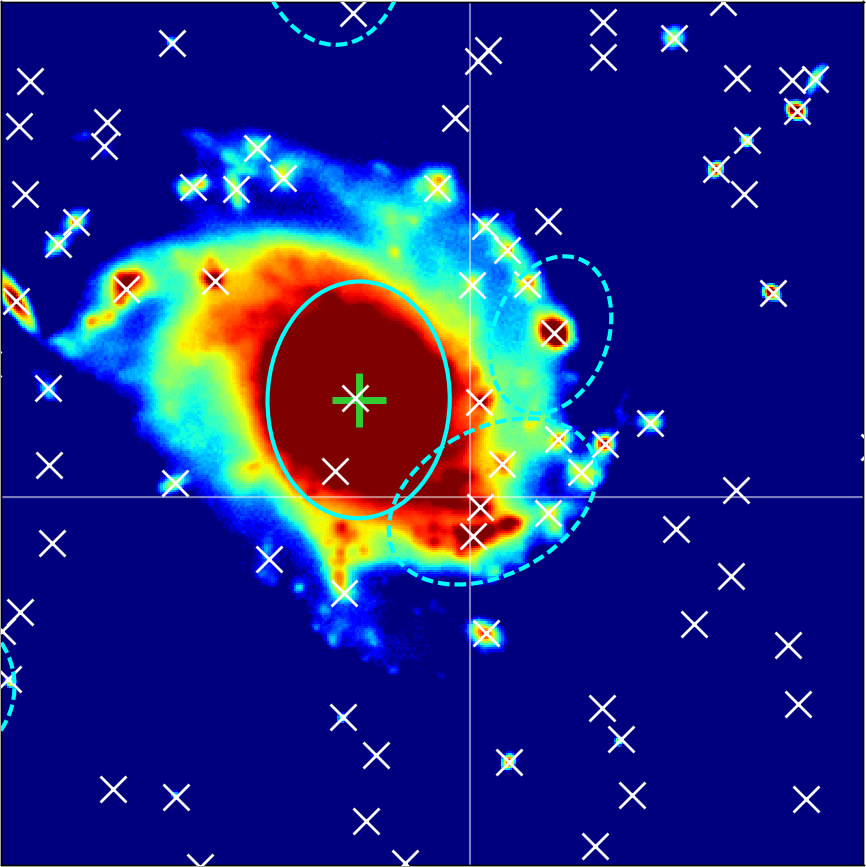}
    \end{minipage}  
    \begin{minipage}[t]{.24\textwidth}
        \centering
        \includegraphics[width=0.95\textwidth]{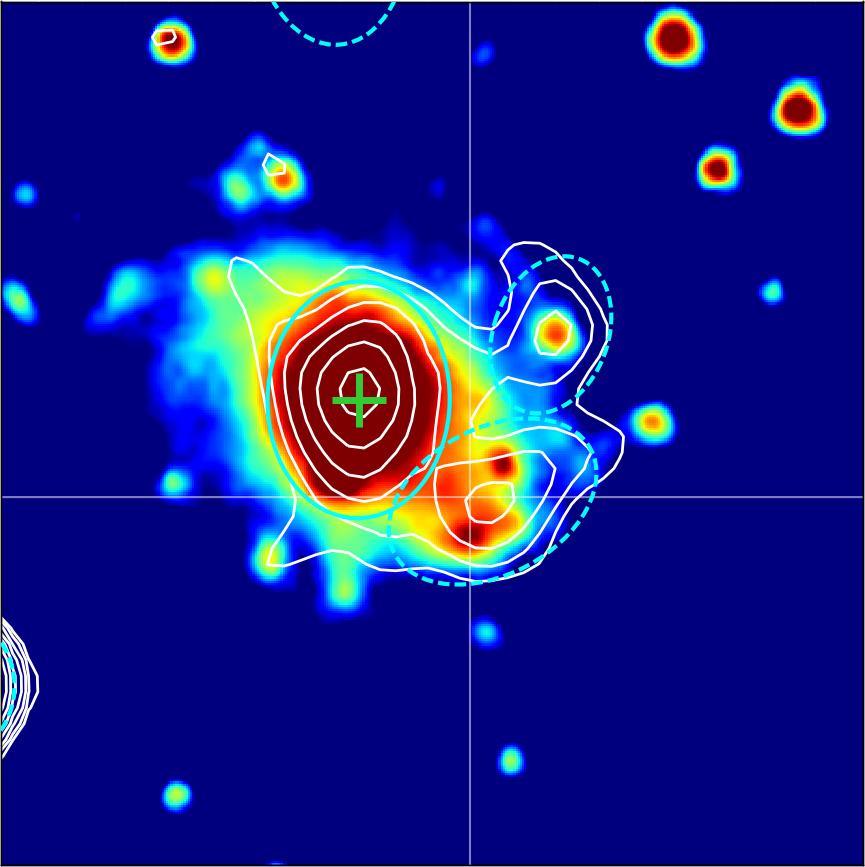}
    \end{minipage}
    \begin{minipage}[t]{.24\textwidth}
        \centering
        \includegraphics[width=0.95\textwidth]{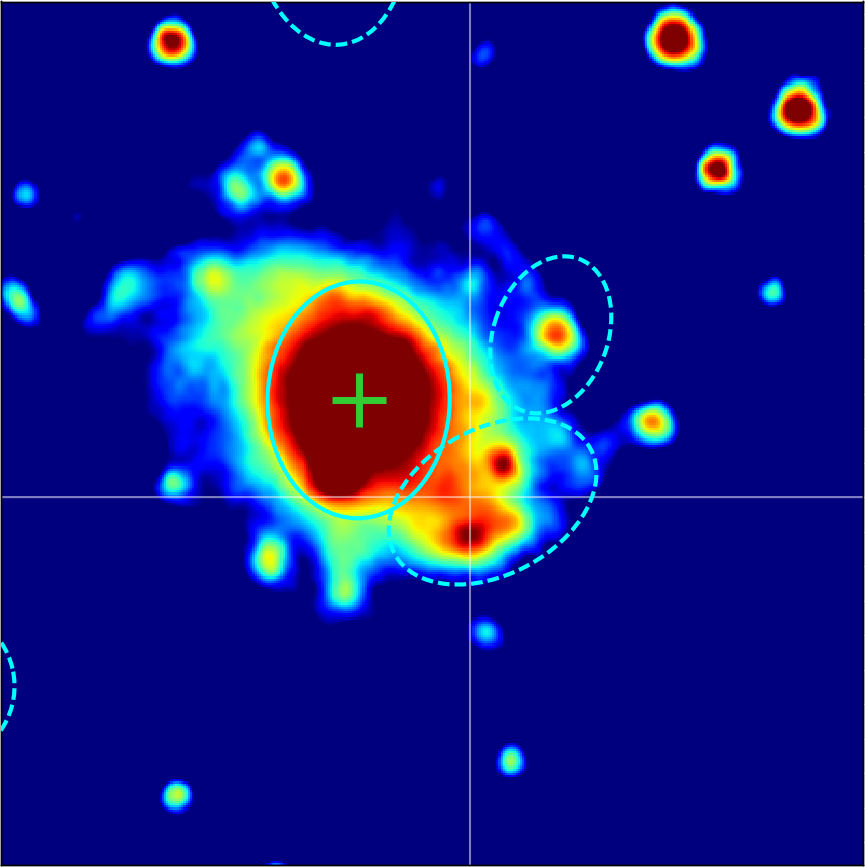}
    \end{minipage}  
    \linebreak
    \linebreak
    \begin{minipage}[t]{.24\textwidth}
        \centering
        \includegraphics[width=0.95\textwidth]{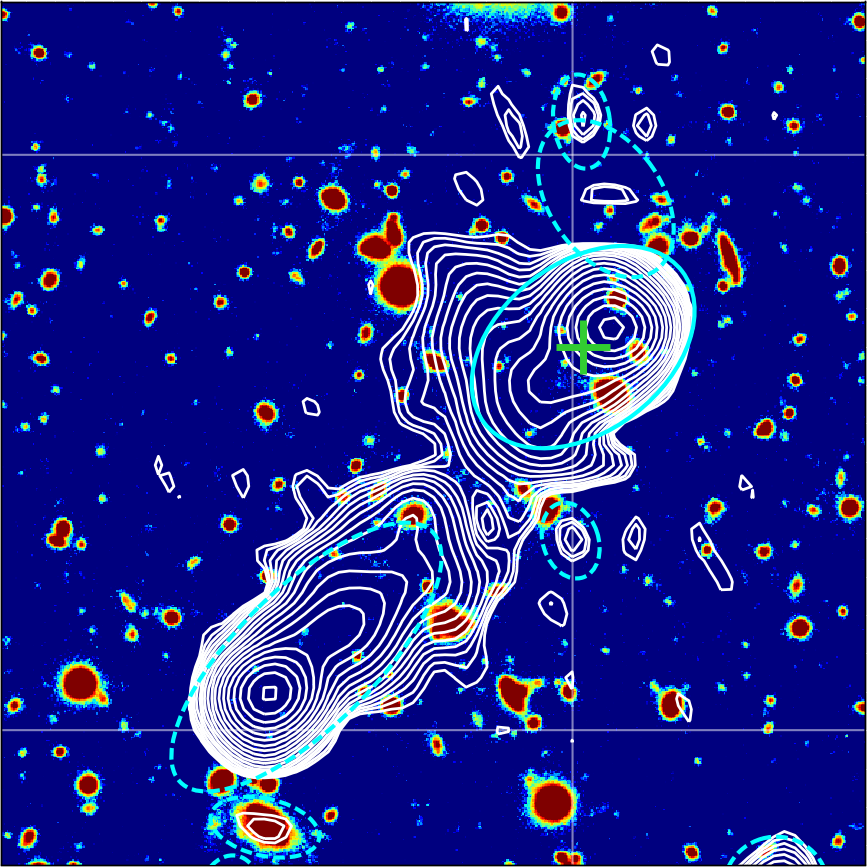}
    \end{minipage}
    \begin{minipage}[t]{.24\textwidth}
        \centering
        \includegraphics[width=0.95\textwidth]{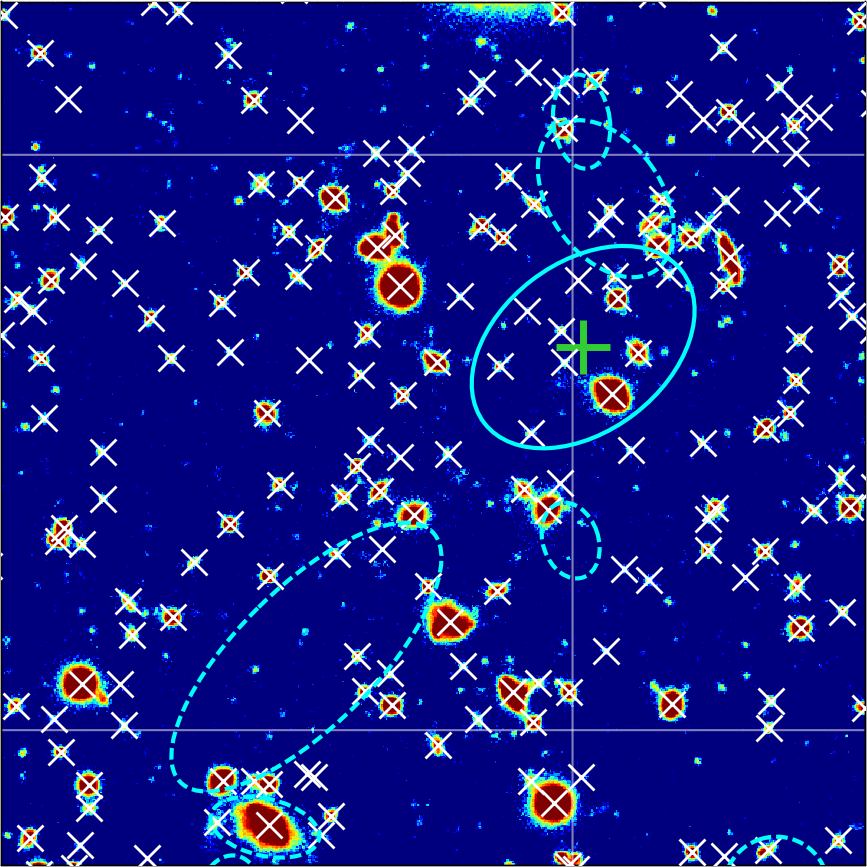}
    \end{minipage}  
    \begin{minipage}[t]{.24\textwidth}
        \centering
        \includegraphics[width=0.95\textwidth]{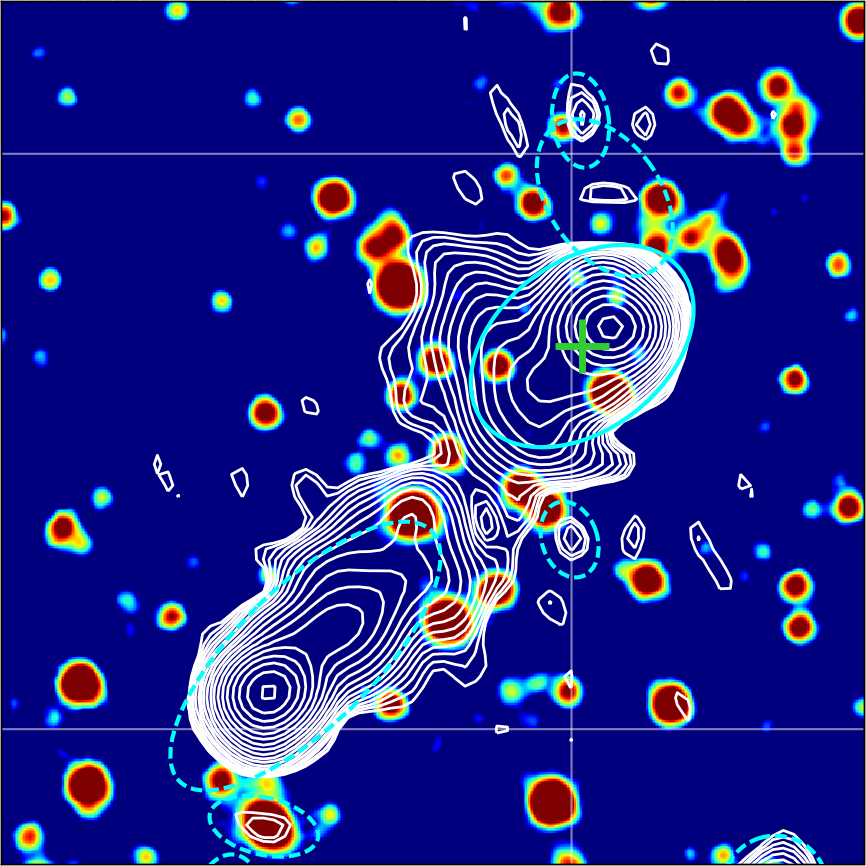}
    \end{minipage}
    \begin{minipage}[t]{.24\textwidth}
        \centering
        \includegraphics[width=0.95\textwidth]{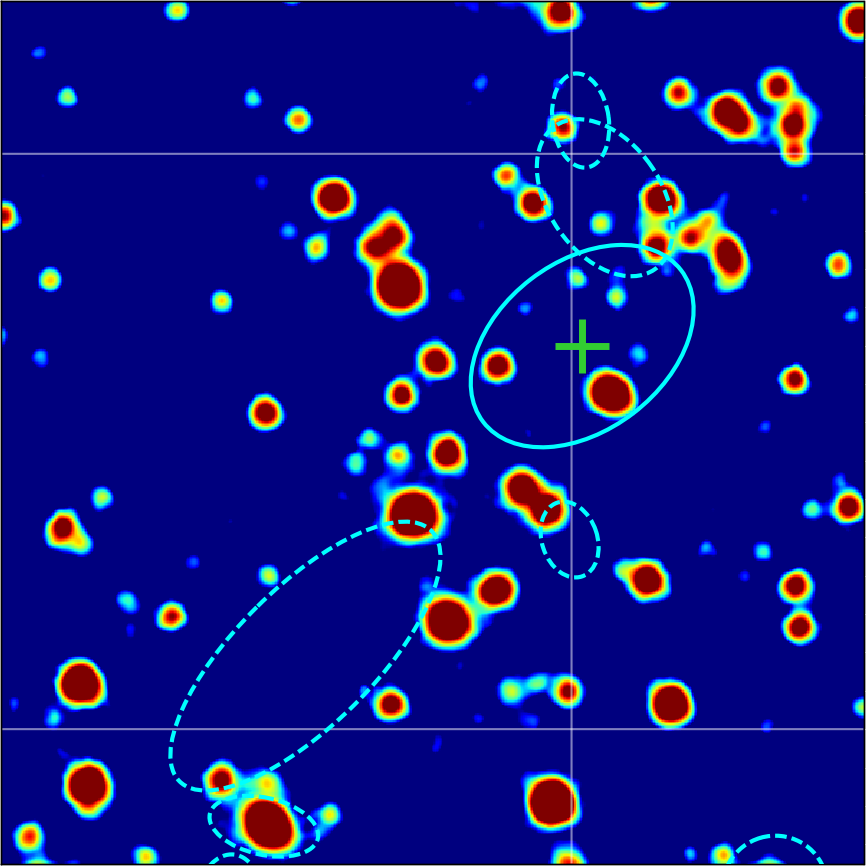}
    \end{minipage}  
    \linebreak
    \linebreak
    \begin{minipage}[t]{.24\textwidth}
        \centering
        \includegraphics[width=0.95\textwidth]{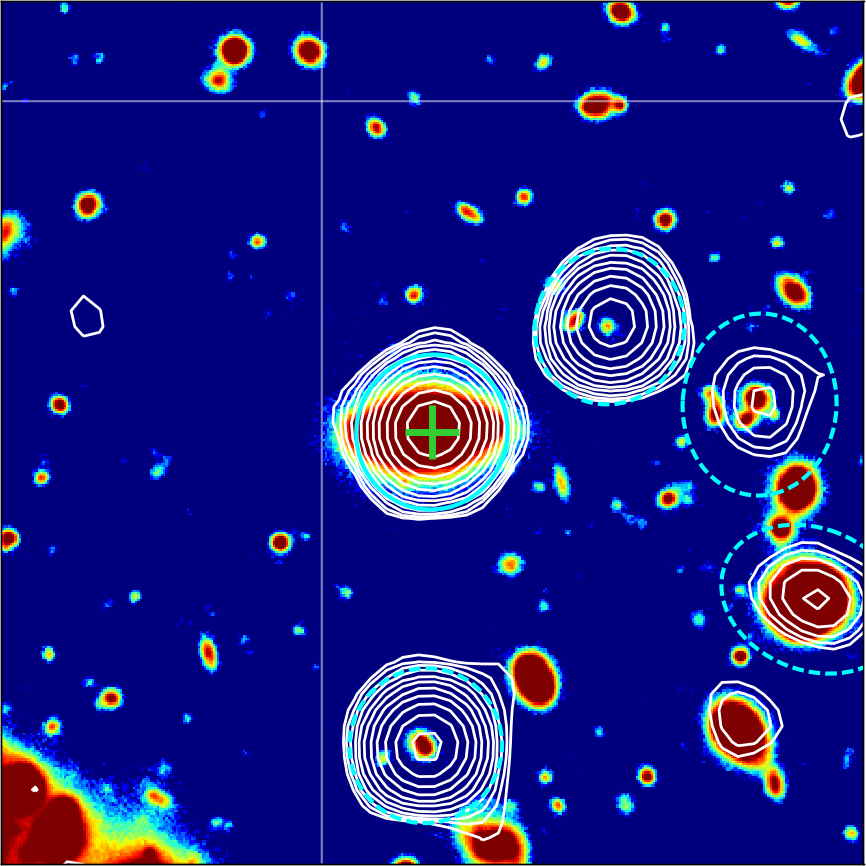}
    \end{minipage}
    \begin{minipage}[t]{.24\textwidth}
        \centering
        \includegraphics[width=0.95\textwidth]{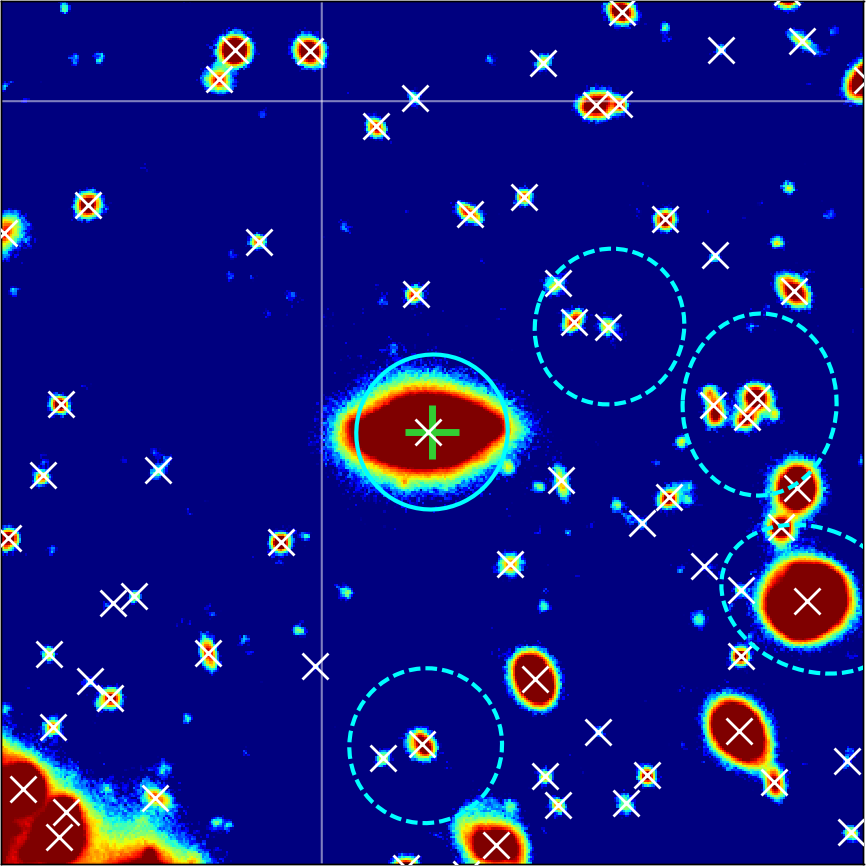}
    \end{minipage}  
    \begin{minipage}[t]{.24\textwidth}
        \centering
        \includegraphics[width=0.95\textwidth]{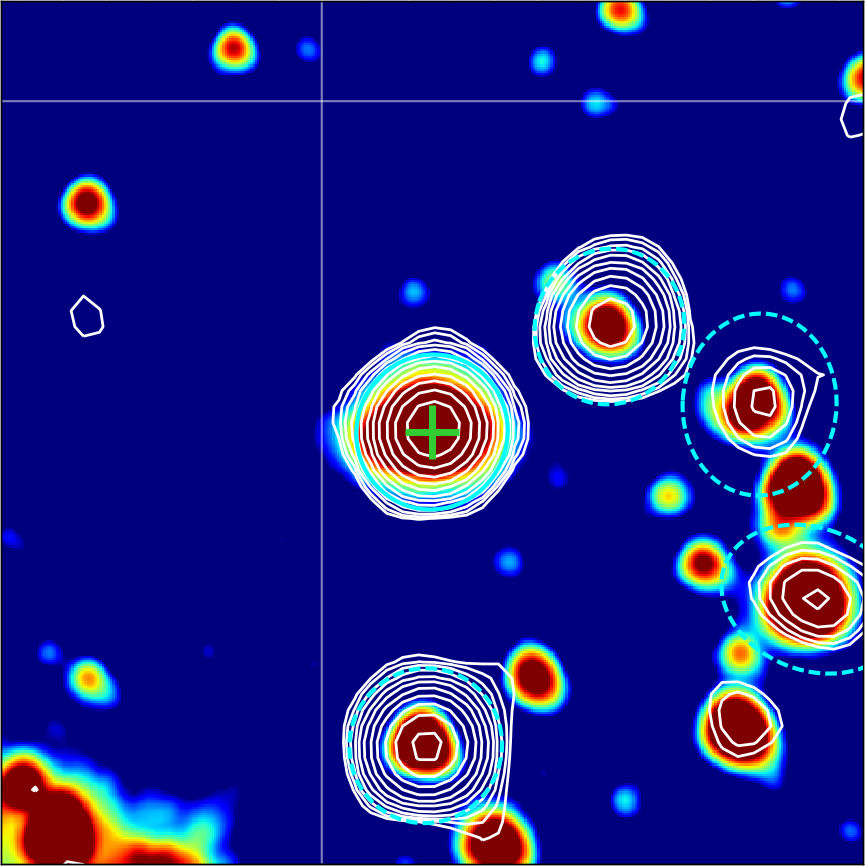}
    \end{minipage}
    \begin{minipage}[t]{.24\textwidth}
        \centering
        \includegraphics[width=0.95\textwidth]{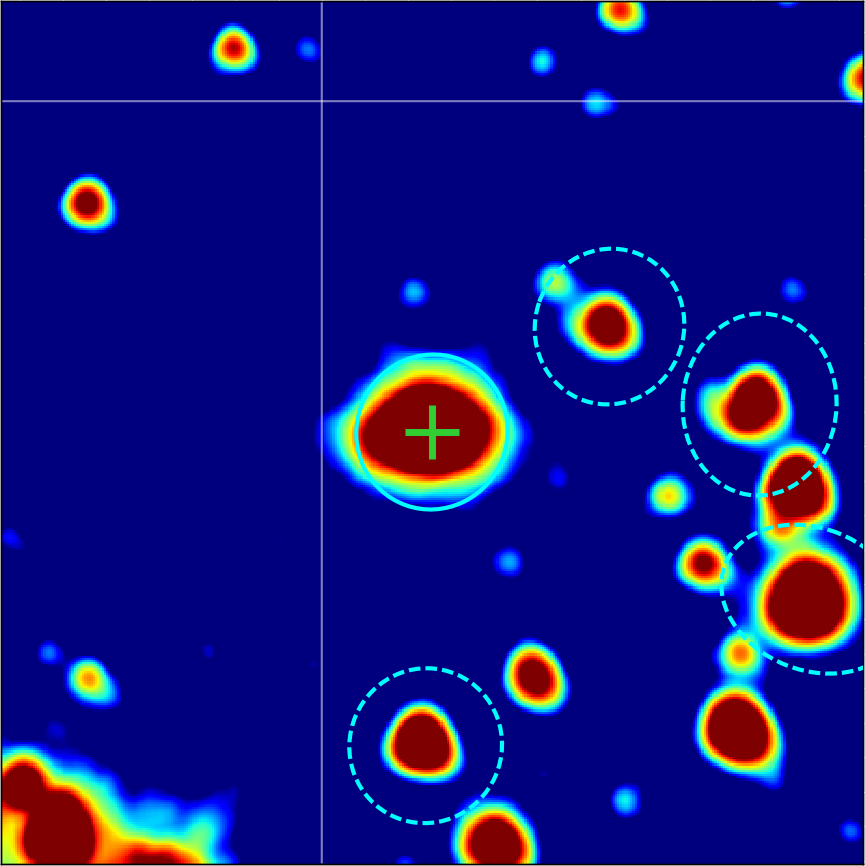}
    \end{minipage}  
    \caption{Images for three sources (upper: a nearby spiral galaxy; middle: a jetted AGN; lower: a compact radio source) which were used in the galaxy zoo platform to help identify host galaxies and associate radio components. Shown are the images available to MIGHTEE zoo participants, namely: i-band imaging background with white radio contour overlays (left); i-band imaging background with white crosses indicating the location of potential K-band selected host galaxies (centre left); 3.6 $\muup$m imaging from Spitzer as the background with white radio contour overlaid (centre right); and 3.6 $\muup$m imaging from {\emph{Spitzer}} as the background only. On each panel a solid cyan ellipse is used to indicate the radio source to be cross-matched and additional radio sources identified by \texttt{PyBDSF} are indicated by cyan dashed ellipses.}
    \label{fig:zooimages}
\end{figure*}

\subsection{MIGHTEE Zoo}
\label{sec:zoo}
In order to visually cross-match the sources which could not be robustly assigned a host using the likelihood ratio method, we follow the method used by the LOFAR surveys collaboration \citep[see][]{Williams2019, Kondapally2021, Hardcastle2023} and set up a Zooniverse\footnote{\url{https://www.zooniverse.org}} project \citep[see e.g.][]{Zooniverse} internal to the MIGHTEE consortium {\citep[see the process described in][]{Hardcastle2023}}. Each source sent to the Zooniverse was visually inspected by at least 5 members of the MIGHTEE collaboration, where they were asked a series of questions. These questions were used to (1) identify additional \texttt{PyBDSF} sources which are associated with the same galaxy; (2) {if possible,} identify a host galaxy; (3) flag the source if necessary to identify cases where e.g. the object was an artefact or the image was too zoomed in to see the full source extent and {(4)} provide a text box to (optionally) describe the source. Example images which were shown in the zoo are presented for three sources in Figure \ref{fig:zooimages}. Four images are created per source: two use $i$-band images from HSC-SSP as the background and the others use 3.6 $\muup$m images from either the {COSMOS Spitzer survey \citep[S-COSMOS;][]{Sanders2007} or the Spitzer Extragalactic Representative Volume Survey \citep[SERVS;][]{Mauduit2012}.}

As outlined in Table \ref{tab:maskingdata}, a total of $\sim$10 000 sources required {visual inspection}, resulting in a need for $\sim$50 000 inspections in order for each source to have at least 5 responses from different people before it could be retired from the zoo. {Using the criteria adopted in \cite{Williams2019}, a match would be considered good if $>3/5$ of the classifications from individuals were in agreement.} A total of $\sim$70 volunteers from the MIGHTEE {consortium} across $\sim$10 countries registered to assist with the cross-matching process. Of this, $\sim$45 people eyeballed over 100 sources, and $\sim$20 people {inspected} over 1 000 sources each. The visual identification process was first run {only on the} COSMOS field in 2023. This allowed core members of the MIGHTEE cross-matching team to investigate results and to identify potential challenges which were encountered by {the} team or improvements to the guidance which could be given. The larger cross-matching effort over the CDFS-DEEP and XMM-LSS fields was then conducted in 2024. 

\begin{figure*}
    \includegraphics[width=\textwidth]{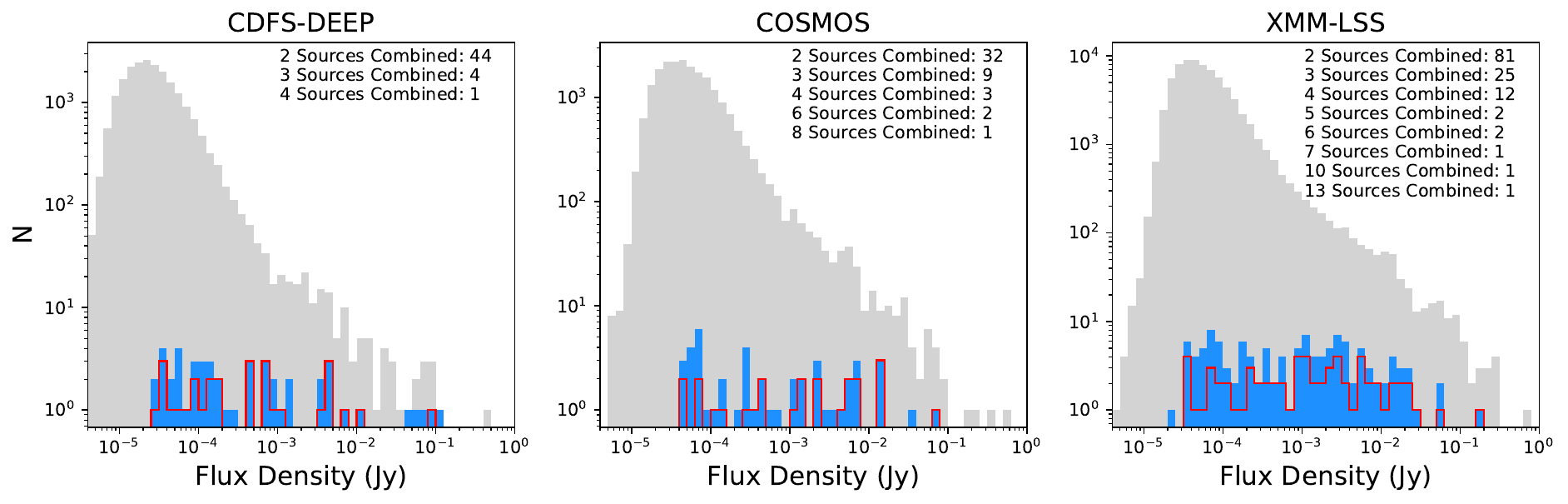}
    \caption{Flux density distribution of sources which were merged together (blue) compared to the final catalogue of sources within the multi-wavelength regions (light grey) for the CDFS-DEEP, COSMOS and XMM-LSS fields (left to right {respectively}). Also shown {are sources which were} checked in the additional steps of Section \ref{sec:cat_checks} (red) and the number of input \texttt{PyBDSF} sources which were combined together into individual output sources from the zoo and given in the top right of each figure.}
    \label{fig:flux_combined_sources}
\end{figure*}

\section{Combined cross-matched Catalogue}
\label{sec:results}
\subsection{Initial combined catalogue from LR and Zoo}
To obtain our cross matched catalogue across each of the three fields we combined the results from the {LR} with those {from visual classification} in the zoo. Sources which were selected by the decision tree to use the {LR} criterion (either as sources or Gaussians) were included in the final catalogue. Next, sources which underwent visual inspection needed to be combined into the catalogue. Following the work of \cite{Williams2019} and \cite{Kondapally2021}, those sources in which 60 per cent of inspectors agreed on the host galaxy {(i.e. 3 out of the 5)}, as well as the association of any \texttt{PyBDSF} sources into the same object, were combined together. Sources in which there was no consensus of a host galaxy from the zoo (including those where there was agreement that multiple \texttt{PyBDSF} sources were combined into a single object) were still included in the final catalogue {with an updated radio position but without a corresponding host RA/Dec}. For those sources in which multiple \texttt{PyBDSF} sources were combined to create a new {radio} source, {new source properties were assigned, namely a flux weighted host position, a new total flux (summed from the individual} \texttt{PyBDSF} sources) and size \citep[see][]{Williams2019}. 

{We note that for those sources where a consensus host galaxy was assigned in the zoo, $\sim$99 per cent of these had $>$80 per cent agreement from visual inspectors and $\sim$77 per cent, $\sim$78 per cent and $\sim 65$ per cent had 100 per cent agreement on the host in CDFS-DEEP, COSMOS and XMM-LSS respectively. Additionally, for those sources where \texttt{PyBDSF} source association was performed in the zoo $\gtrsim$ 70 per cent of these had at least 4/5 visual inspectors agree, and $\sim 15-20$ per cent had all five inspectors agree (though see additional classifications in Section \ref{sec:cat_checks}). These suggest the associations are broadly robust and any mis-classifications from the zoo would be a small fraction of the full catalogue.}

\begin{figure*}
    \includegraphics[width=\textwidth]{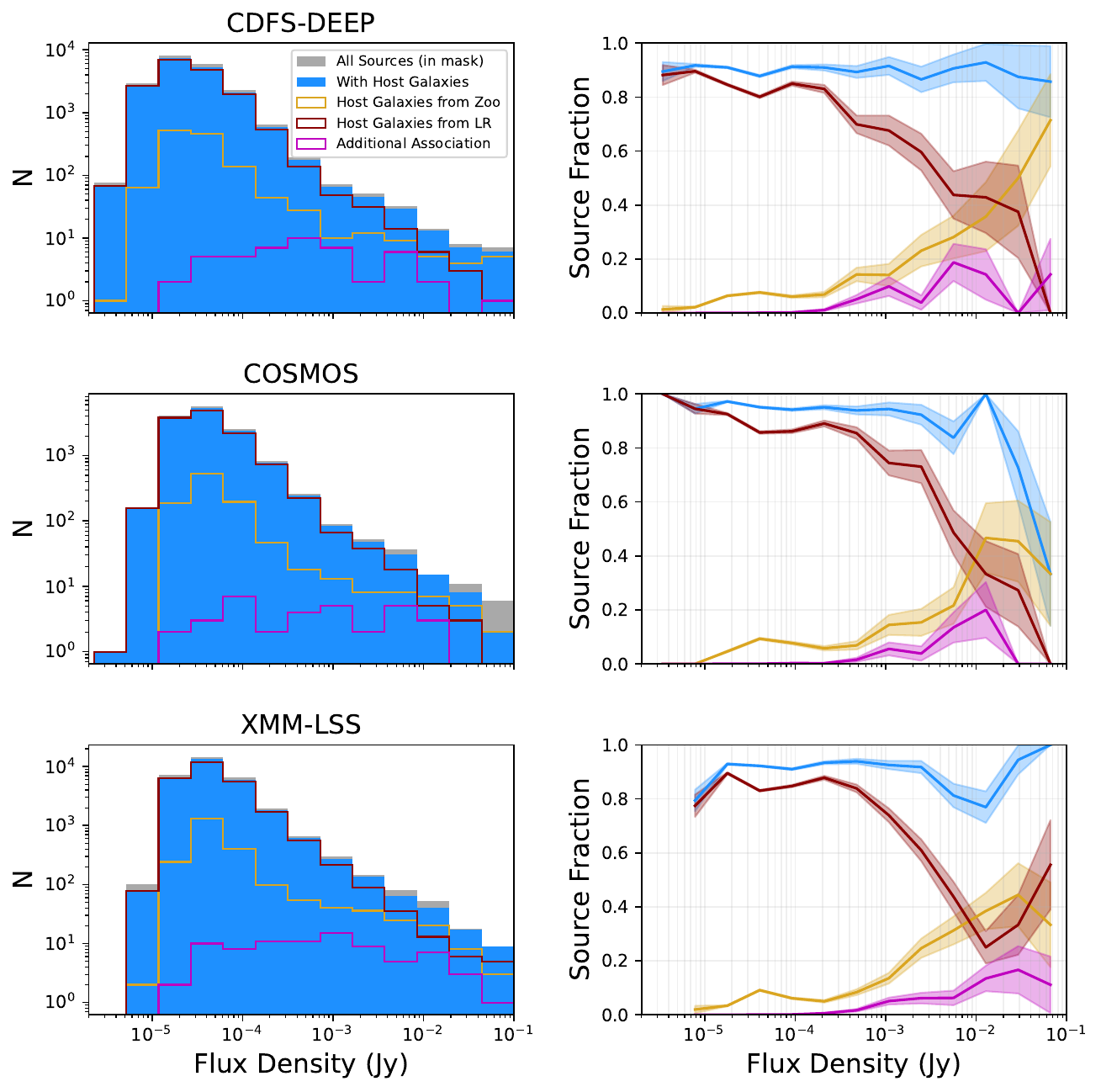}
    \caption{Upper: the flux density distribution of sources in the CDFS-DEEP, COSMOS and XMM-LSS fields (left to right {respectively}) for all sources which are within the unmasked, multi-wavelength regions (grey) and split into those which have a host galaxy associated with the source (blue) and then subset into those sources which have host galaxies associated from: the zoo (gold), the LR method (red), and the additional steps of Section \ref{sec:cat_checks} (pick). Lower: the total fraction of sources in the masked region which have host galaxies associated, and the fraction which come from the zoo and LR method, using the same colours as for the histograms. {The associated errors provided are from the binomial errors.}}
    \label{fig:flux_with_without_hosts}
\end{figure*}

\begin{figure*}
    \includegraphics[width=\textwidth]{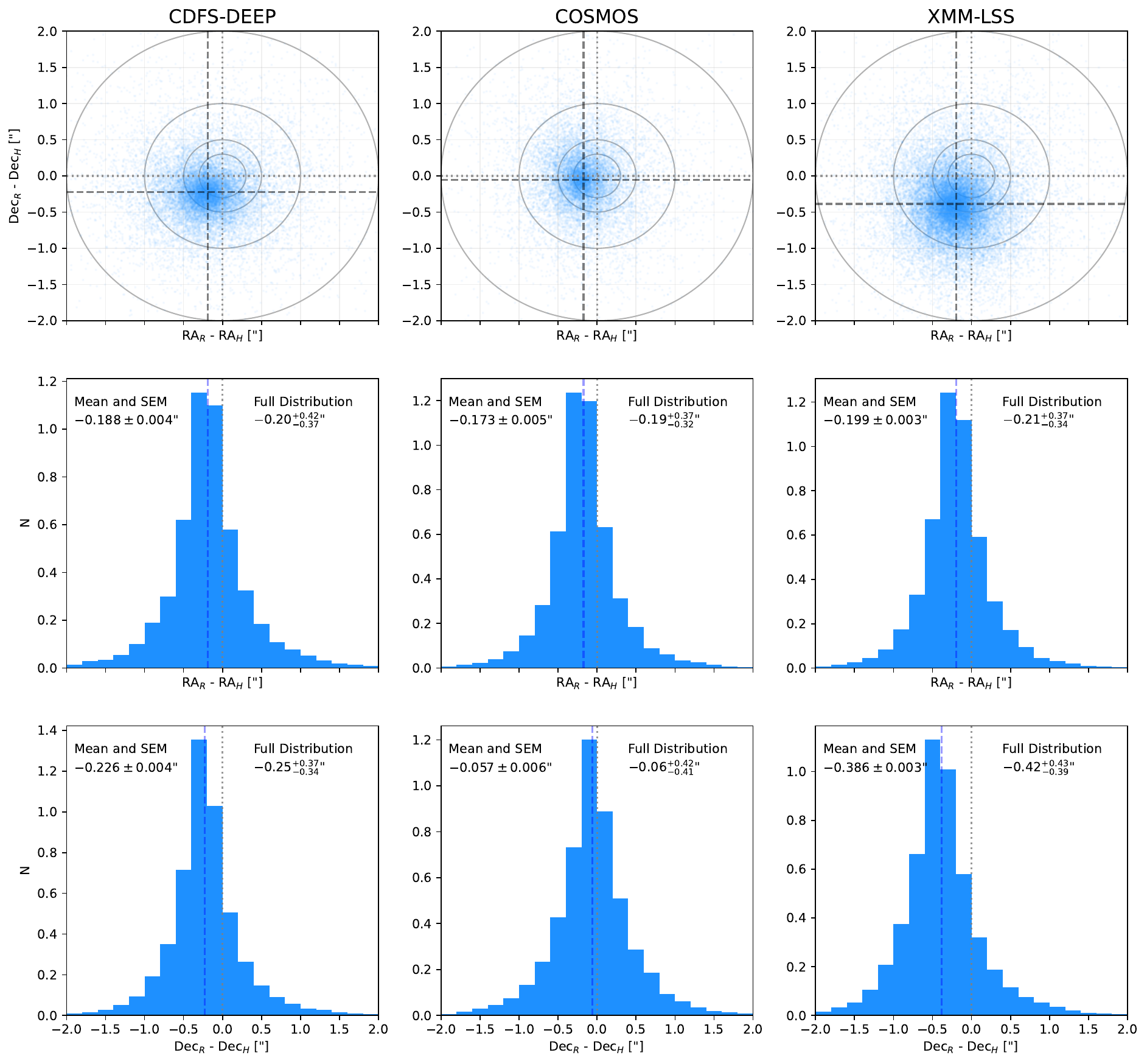}
    \caption{Comparisons of the positions of the radio sources to {those} of the host multi-wavelength {galaxies} (where available) from the `RA\_host' and `Dec\_host' columns (see Appendix \ref{app:catalogue}). Shown are the positional offsets for the CDFS-DEEP, COSMOS and XMM-LSS fields (left to right) shown as: (i) scatter of the offsets (top row), (ii) histogram of the RA offset (middle row) and (iii) histogram of the Dec offset (bottom row). Each offset is defined as the difference in the RA or Dec values (radio position - host position), given in arcsec. The mean offset and standard error on the mean (SEM) is recorded in the the lower two rows (top left) as is the median and errors from the 16th and 84th percentiles (top right). Angular radii of 0.3 {arcsec}, 0.5 {arcsec}, 1 {arcsec} \ and 2 {arcsec} \ are shown in the top panels. The mean offset is shown as a black dashed line, as well as a grey dotted line to indicate no positional offset (for reference only).  We note that a small fraction of sources ($\lesssim$1 per cent per field) have positional offsets larger than the 2 {arcsec} \ limits shown.}
    \label{fig:host_offsets}
\end{figure*}

\subsection{Additional Checks on the Catalogues}
\label{sec:cat_checks}
Finally, additional checks were performed on this merged catalogue to ensure the quality of the catalogue. The checks performed identified a number of corrections which needed to be made to finalise the catalogue. This included identifying:

\begin{enumerate}
    \item Sources which were in the original LR catalogue, but had also been incorporated within the zoo catalogue and thus had been left as duplicate sources within the catalogues
    \item Sources flagged {as too zoomed in} by $\geq$60 per cent of inspectors within the Zoo framework (i.e. `Zoom\_prob'$\geq$0.6, as described in Appendix \ref{app:catalogue}).
    \item Sources where the host galaxy assigned was the same as another within the catalogue
    \item Sources where either the radio positions or host galaxy positions were close to one another ($\leq$1 {arcsec})
    \item Sources which consisted of a component which was included in multiple output zoo sources
\end{enumerate}

\noindent Of these criteria, the last {two} typically represented sources which were a combination of (i) sources which were split into Gaussian components, but where multiple of these Gaussians were assigned the same host galaxy or (ii) sources which had been output from the zoo but where there appeared to be some disagreement between classifiers on how many initial radio \texttt{PyBDSF} sources to include within the full extent of the source. 

{Sources} which met any of these {five} {criteria were} {additionally} eyeballed by the lead author in order to make a decision on the source (i.e. which \texttt{PyBDSF} sources to combine, if sources should be merged and if a host galaxy should be assigned). This led to $\sim50-300$ sources being investigated per field. Where any changes to the sources in the catalogues were required, these were applied by either removing or merging sources in the catalogue. These sources were flagged within the catalogue through the `match\_method' column, see Appendix \ref{app:catalogue}.

The flux distribution of all sources in this final catalogue is shown in Figures \ref{fig:flux_combined_sources} and \ref{fig:flux_with_without_hosts} for those sources which are within the multi-wavelength regions. In Figure \ref{fig:flux_combined_sources}, the flux density distribution of those sources which were combined together from the initial \texttt{PyBDSF} catalogue either through the Zoo or through the associations outlined in this section are presented, and the number of components which were combined together to make these sources is presented. In Figure \ref{fig:flux_with_without_hosts}, the flux {density} distribution is {instead} shown for all sources and shown separately as those sources with (a) a host assigned and (b) sub-categorised based on the method used to assign a host. 

We present the distribution of the positional offsets between the radio positions and the host galaxy positions in Figure \ref{fig:host_offsets} per field. For each field, a non-zero mean offset in both RA and Dec is {found}. These mean offsets are typically $\lesssim$0.2 {arcsec}, except for the XMM-LSS field, where the {absolute} mean declination offset is closer of 0.4 {arcsec}. However, all mean offsets in both RA and Dec are smaller than the pixel size of the MIGHTEE images (1.1 {arcsec}), and thus are within expected values.

\begin{figure}
    \centering
    \includegraphics[width=\linewidth]{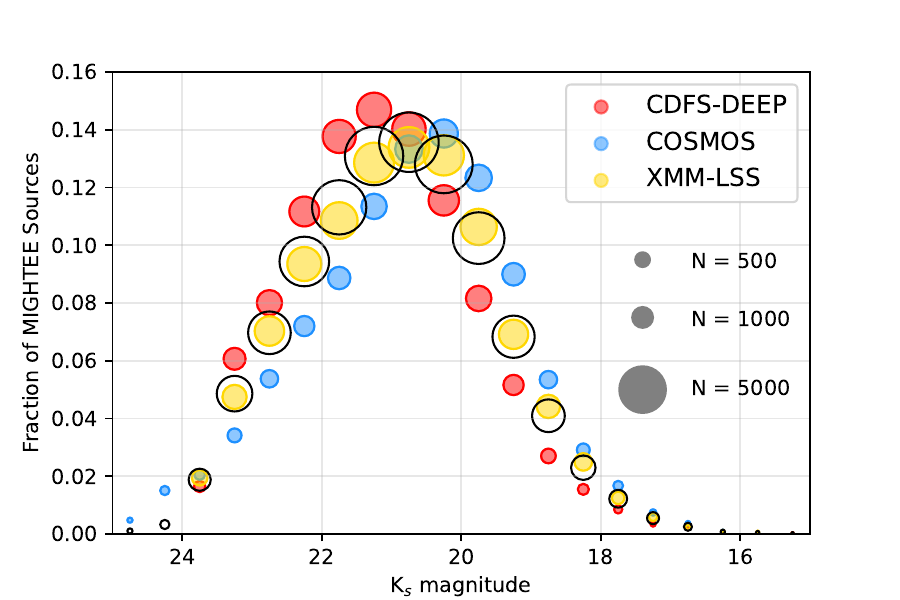}
    \includegraphics[width=\linewidth]{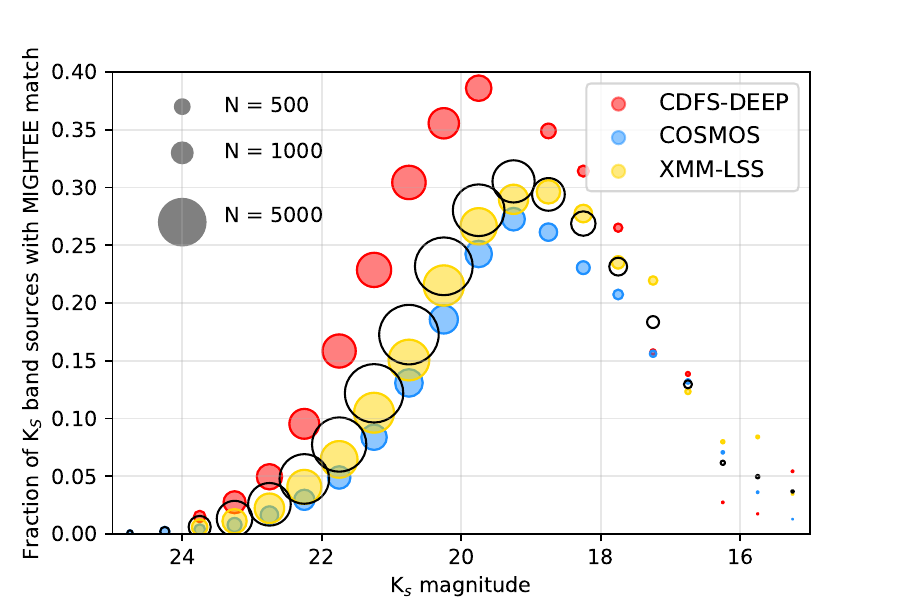}
    \caption{{Comparison of the $K_{s}$ band magnitude distribution of sources which are hosting one of the MIGHTEE sources. \textit{Upper:} Magnitude distribution of our radio sources which have a $K_{s}$ band magnitude as a fraction of all MIGHTEE sources which have a host galaxy assigned. \textit{Lower:} The fraction of all $K_{s}$ band sources with an associated MIGHTEE source. Shown in each panel are the CDFS-DEEP (red), COSMOS (blue) and XMM-LSS (yellow) fields and their size indicates the number of radio sources contributing (see the grey circles for an indication of numbers). The black open circles represent the combined values across the three fields. }}
    \label{fig:mag_frac}
\end{figure}

{We additionally present the comparison of the fraction of sources which have a MIGHTEE source as a function of $K_s$ band magnitude in Figure \ref{fig:mag_frac}, in a similar vein to Fig. 7 of \cite{Kondapally2021}. This presents both (i) the fraction of the MIGHTEE radio sources within the multi-wavelength regions and (ii) the fraction of the full $K_s$ catalogue of sources which overlap with the radio data. We present this both for each field individually, alongside the total fraction across the three fields. Figure \ref{fig:mag_frac} demonstrates that the CDFS-DEEP field detects fainter radio sources than the other two fields, with the fraction of MIGHTEE sources peaking at fainter $K_s$ band magnitudes than the other two fields. The $K_s$ band sources which have the largest fraction of MIGHTEE counterparts have a magnitude of $\sim$19-20, with between 25-40 per cent of $K_s$ sources in this magnitude bin containing a radio source. However, for the radio population themselves, these peak closer to fainter $K_s$ band magnitudes at $\sim$21.    }

\subsection{Redshifts}
\label{sec:redshifts}

As outlined in Section \ref{sec:intro}, {redshifts are} imperative to studies of galaxies {in order} to accurately place galaxies within the context of cosmic history. In this section we present the combination of photometric and spectroscopic redshifts for the radio sources.

\subsubsection{Photometric Redshifts}
To assign photometric redshifts to the combined radio source catalogue we use the catalogues adopted in Stylianou et al. (in prep). As discussed, at the time of commencing the cross-matching process, a previous version of $K_s$ catalogues were used to assign host galaxies. {In COSMOS, the UltraVISTA data were updated to DR6\footnote{\url{https://doi.org/10.18727/archive/52}} in the catalogues of Stylianou et al. {(in prep)} and in the XMM-LSS and CDFS fields, the catalogues were also updated between the source cross-matching and redshift association process. For all fields, there were also small differences in the masking which were applied to these catalogues,} this is recorded in the catalogue, see Appendix \ref{app:catalogue}. Therefore, we match the host galaxy positions to the positions in the updated catalogues of Stylianou et al. using a 1 {arcsec} match radius. These updated source positions are provided in the catalogues released with this work, as outlined in Appendix \ref{app:catalogue}. 

For the catalogues of Stylianou et al. {(in prep)}, there are two {origins} for the photometric redshifts, depending on the field. In the COSMOS and XMM-LSS fields, the redshifts are predominantly from the compilation of photometric redshifts from \cite{Hatfield2022} {based on 10 photometric bands: four VISTA filters \citep[from][]{McCracken2012, Jarvis2013}, five HSC filters and the u-band filter from the Canada-France-Hawaii Telescope Large Area U-band Deep Survey, CLAUDS \citep{CLAUDS}. \cite{Hatfield2022} uses a Hierarchical Bayesian model \citep[see][for methodology]{Duncan2018} to combine photometric redshift estimates from template fitting using the code \texttt{LePHARE} \citep[see][]{LePHARE2, LePHARE1}} with machine learning techniques from the code GPz \citep{Almosallam2016a, Almosallam2016b}. In the CDFS-DEEP field, however, the catalogues of \cite{Hatfield2022} were not available and so Stylianou et al. {(in prep)} carried out template fitting of the source fluxes using the code \texttt{LePHARE}. {For each source, the photometric redshifts were obtained by matching the observed photometry with a set of redshifted galaxy SED templates (with and without the incusion of AGN) and selecting the best-fitting redshift through $\chi^2$ minimisation. The template set adopted here was the COSMOS library included in \texttt{LePHARE}, which contains 32 SEDs from the \cite{Ilbert2009} COSMOS photo-z work. The templates were allowed to vary with dust attenuation using the \cite{Calzetti2000} extinction law, and the fitting used only the available photometric bands for each galaxy, with a minimum flux-error floor of 5 per cent.} \texttt{LePHARE} was additionally run for those sources in the COSMOS and XMM-LSS fields for sources which were not in the catalogues of \cite{Hatfield2022}. {For sources with \texttt{LePHARE} only templates we select either the galaxy only best-fitting redshift or the redshifts where AGN templates were also included, depending on which redshift minimised the $\chi^2$ value.} 

For each of the redshifts available, an associated probability distribution function (pdf) was available and therefore we record several properties of the pdf for the associated radio source to quantify the uncertainties in the redshift distribution. These properties are: the peak of the photometric redshift, the 1$\sigma$ confidence interval around the peak redshift, and the values of the 50th, 16th and 84th percentiles of the pdf. Additionally, for the COSMOS field, we match to sources within the Physics of the Accelerating Universe (PAU) survey catalogue \citep{PAU}. If available, the PAU redshifts were adopted to supersede those from \cite{Hatfield2022}. In total, {18 380} sources in the CDFS-DEEP field had photometric redshifts available {(corresponding to $\sim$90 per cent of sources in the masked region)}, increasing to  {91 per cent (28 493 sources)} in the XMM-LSS field and 95 per cent ({13 117 sources}) in the COSMOS field.

{As discussed in \cite{Hatfield2022}, the combined method of template fitting and machine learning can help improve the accuracy of the redshifts, and so this may lead to more biases within the CDFS redshifts. In general, with photometric redshifts, a lack of availability of spectroscopic redshifts and samples which cover the broad distribution of galaxies (especially at high $z$) will affect the accuracy of photometric redshifts. Such photometric redshifts will be improved in these fields in the future thanks to additional deep ancillary data provided by the Large Synoptic Survey Telescope \citep[LSST;][]{LSST} and Euclid \citep{Euclid}. }

\begin{figure*}
	\includegraphics[width=18cm]{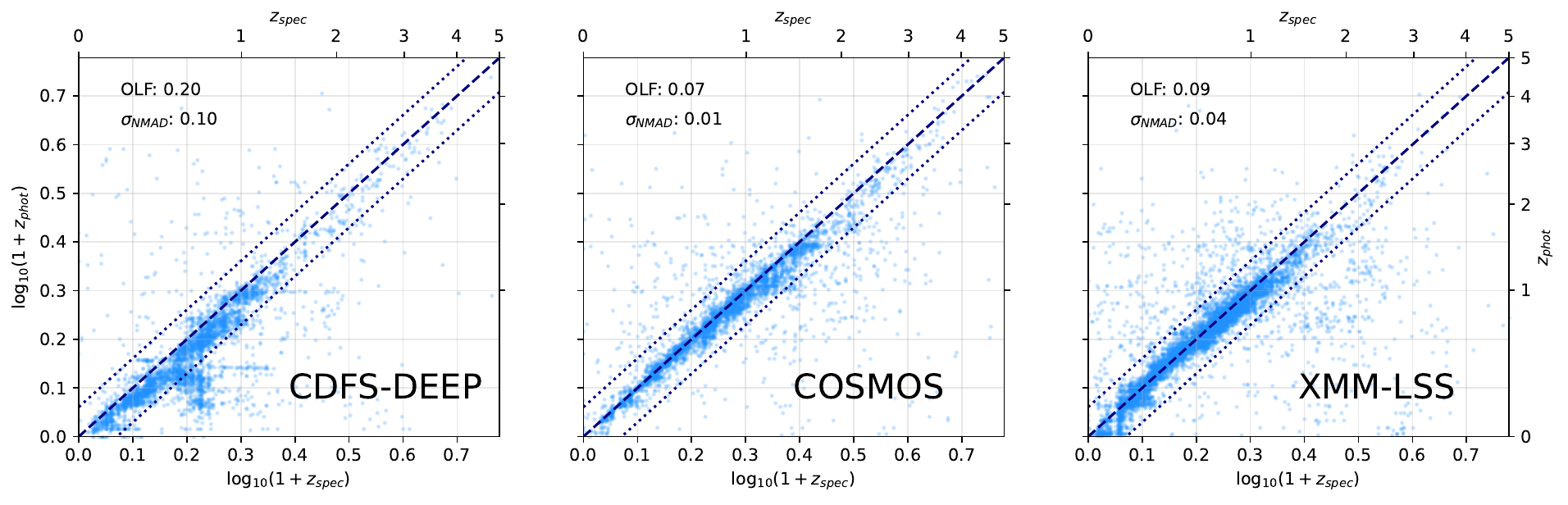}
	\caption{Comparison of the spectroscopic redshifts (x-axis) to the corresponding photometric redshift (y-axis, given as the peak in the redshift pdf) for sources in the CDFS-DEEP (left), COSMOS (centre) and XMM-LSS (right) fields. Also shown to guide the eye are the 1-to-1 line (dashed) and dotted lines which indicate the regime for the outlier fraction (OLF), using Equation \ref{eq:olf} (dotted).}
	\label{fig:specz_photz}
\end{figure*}

\begin{figure*}
    \includegraphics[width=18cm]{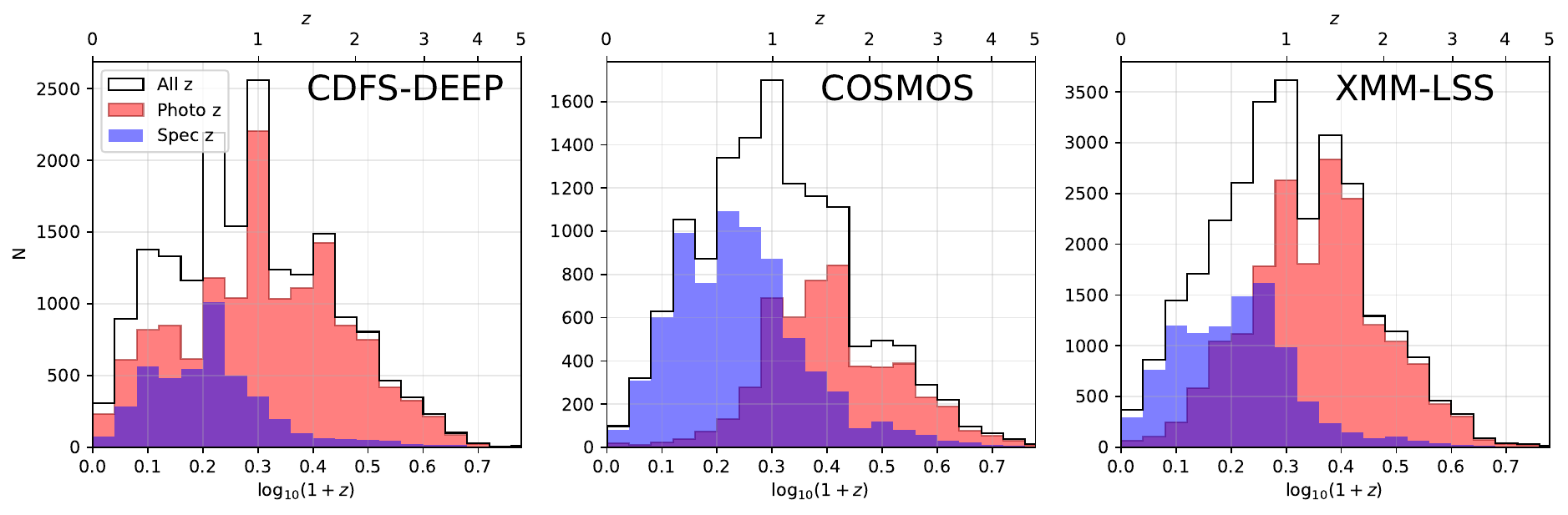}
    \caption{Histogram of the best redshifts for all available sources split into the contributions for photometric redshifts (red) and spectroscopic redshifts (blue) for the CDFS-DEEP (left), COSMOS (centre) and XMM-LSS (right) fields. The combined distribution from the photometric and spectroscopic redshifts are shown in the black outline. }
    \label{fig:zdist}
\end{figure*}

\subsubsection{Spectroscopic Redshifts}
\label{sec:specz}
{Due} to the fact that these fields are well {studied} across the electromagnetic spectrum, there also exist numerous spectroscopic surveys which provide more accurate redshifts across the fields. We use a compilation of redshifts within the three deep fields\footnote{Using the version from {20th March 2025}} \citep[see e.g.][]{2022specz, 2026specz} to identify spectroscopic counterparts for our radio sources, using a 1 {arcsec} match radius. These spectroscopic redshifts are a compilation from a variety of spectroscopic catalogues including {(but not limited to) those} from {Dark Energy Spectroscopic Instrument survey \citep[DESI][]{DESI2025}, Galaxy And Mass Assembly survey \citep[GAMA;][]{Driver2022} and Sloan Digital Sky Survey \citep[SDSS;][]{Ahumada2020}} which are outlined in Appendix \ref{app:catalogue}. If a spectroscopic redshift was available, this was adopted as the best redshift for the source. 

Across the three fields the spectroscopic coverage varies, leading to a total of {4 393} {sources in with spectroscopic redshifts in CDFS-DEEP}, {7 256} sources in COSMOS and {9 871} in XMM-LSS. This corresponds to {21.5 per cent, 52.6 per cent and 31.4} per cent of the radio catalogues within the unmasked regions of the CDFS-DEEP, COSMOS and XMM-LSS fields. A comparison of the spectroscopic redshifts to the corresponding photometric redshift {(taken as the peak of the pdf)} is presented in Figure \ref{fig:specz_photz}. We also provide values for two quantities which are commonly used to assess the quality of the photometric redshifts: the normalised median absolute deviation ($\sigma_{\textrm{NMAD}}$) and outlier fraction (OLF). $\sigma_{\textrm{NMAD}}$ is given by:

\begin{equation} 
\sigma_{\textrm{NMAD}}= 1.48 \times \textrm{median}\left(\frac{|\Delta z|}{1 + z_s}\right)
\label{eq:nmad}
\end{equation}

\noindent see \cite{Hoaglin1983}. Here $\Delta z$ is the difference between the spectroscopic ($z_s$) and photometric ($z_p$) redshifts and $\sigma_{\textrm{NMAD}}$ quantifies the spread in the data between the spectroscopic and photometric redshifts. The OLF is instead the fraction of sources which satisfy:

\begin{equation}
    |\Delta z| > 0.15(1+z_s),
    \label{eq:olf}
\end{equation}

\noindent see e.g. \cite{Dahlen2013, Duncan2021} and \cite{Hatfield2022} and the OLF quantifies the fraction of sources with significant deviations (of 15 per cent) between the photometric and spectroscopic redshifts.

Of the fields, CDFS-DEEP has the highest OLF at 0.20 and $\sigma_{\textrm{NMAD}}$ of 0.10, with the COSMOS and XMM-LSS fields having lower OLFs (0.07 and 0.09 respectively) and $\sigma_{\textrm{NMAD}}$ (0.01 and 0.04 respectively). Given that the redshifts for the CDFS-DEEP field are from template fitting alone, whereas the other fields use machine learning techniques to improve the redshifts, this larger OLF and $\sigma_{\textrm{NMAD}}$ are expected {and this can also explain the small offset in redshifts from the 1-to-1 line (which is more apparent due to the log scaling on Figure \ref{fig:specz_photz})}. Furthermore, whilst the values of the OLFs in the COSMOS and XMM-LSS fields are larger than for the full population of galaxies found in \cite{Hatfield2022} (closer to $\lesssim$4 per cent), this is expected given the challenges {of} accurately measuring photometric redshifts from SED fitting when there is a significant population of AGN in the sample \citep[see e.g. discussion in ][]{Duncan2021}. We note that in the CDFS-DEEP field, the spectroscopic redshifts appear systematically higher than the photometric redshifts, with an apparent excess of sources in the CDFS-DEEP field with $z_{\textrm{spec}}$$\sim$0.6-0.7 which have lower photometric redshifts (see Figure \ref{fig:specz_photz}). Users of the catalogues should be aware of {this.}

In total, $\sim$90 per cent of sources in the masked CDFS-DEEP field, $\sim$91 per cent in XMM-LSS field and $\sim$95 per cent in COSMOS have a corresponding redshift associated with them (either spectroscopic or photometric). We present the distribution of the best redshift associated with each source (where available) in Figure \ref{fig:zdist}, separated into those with spectroscopic or photometric redshifts. This demonstrates that the majority of sources for $z \lesssim 1$ in the COSMOS field have spectroscopic redshifts. The COSMOS field has a cumulative fraction of $>80$ per cent of sources with spectroscopic redshifts below $z\sim$1, whereas the cumulative fraction of sources with spectroscopic redshifts below $z\sim$1 is closer to $\sim$55 per cent and $\sim$35 per cent in the XMM-LSS and CDFS-DEEP fields respectively. The fractions of sources with spectroscopic redshifts will increase with the commencement of spectroscopic surveys with telescopes such as the 4-m Multi-Object Spectroscopic Telescope \citep[4MOST;][]{4MOST} and the Multi-Object Optical and Near-infrared Spectrograph \citep[MOONS;][]{MOONS} including, most notably, the Optical Radio Continuum and H{\sc i} Deep Spectroscopic Survey \citep[ORCHIDSS;][]{ORCHIDSS} which will specifically follow up MIGHTEE continuum sources with spectroscopy as part of the Wide-Area VISTA Extragalactic Survey \citep[WAVES;][]{WAVES}. We note that in both Figures \ref{fig:specz_photz} and \ref{fig:zdist} we plot only redshifts up to $z=5$. We note that a small number of sources may have photometric redshifts recorded at $z>5$. {Although} some have been confirmed with spectroscopy \citep[e.g.][]{Varadaraj2026}, we urge caution and {note that these sources} often have a second lower redshift peak in the pdf. \\

\noindent {A summary of the number of sources with redshifts can be found in Table \ref{tab:Nmatches}.}

\begin{table*}
  \centering
\caption{{Details of the number of radio sources over the restricted multi-wavelength regions (see Figure \ref{fig:masking}) before and after host galaxy association. Also listed are the number of sources post host association with redshifts assigned, and the corresponding subsets of these {where the final redshifts (\texttt{z\_best\_final}) are from the} photometric ($z_{\textrm{phot}}$), spectroscopic ($z_{\textrm{spec}}$) and, for the COSMOS field, with redshifts from the Pau survey ($z_{\textrm{Pau}}$). For the subset of sources with redshifts we indicate the fraction of the host associated catalogue this represents.}}
\begin{tabular}{l|rrrr}
\hline
 & CDFS & COSMOS & XMMLSS & Total \\ \hline \hline 
N$_s$ (pre host association) &20 335 & 13 781 & 31 491 &  65 607 \\  
N$_s$ (post host association) &20 458 & 13 782 & 31 412 &  65 652 \\  \hline
N$_s$ With $z$ &18 380 (89.8 {per cent}) & 13 117 (95.2 {per cent}) & 28 493 (90.7 {per cent}) &  59 990 (91.4 {per cent}) \\  
N$_s$ With $z_{\textrm{phot}}$ &13 987 (68.4 {per cent}) & 5 224 (37.9 {per cent}) & 18 622 (59.3 {per cent}) &  37 833 (57.6 {per cent}) \\ 
N$_s$ With $z_{\textrm{Pau}}$ &0 (0.0 {per cent}) & 637 (4.6 {per cent}) & 0 (0.0 {per cent}) &  637 (1.0 {per cent}) \\ 
N$_s$ With $z_{\textrm{spec}}$ &4 393 (21.5 {per cent}) & 7 256 (52.6 {per cent}) & 9 871 (31.4 {per cent}) &  21 520 (32.8 {per cent}) \\ \hline 
\end{tabular}
\label{tab:Nmatches}
\end{table*}

\subsection{Final catalogue of radio sources}
In the final catalogue we combine together those sources from within the multi-wavelength regions (regardless of whether a host galaxy was identified) along with the sources which were in the masked regions {so that} no host was ever sought either through the LR technique or the MIGHTEE zoo\footnote{{We note that for a few sources split into Gaussians, their individual flux density could now be below the nominal 5$\sigma$ individual detection limit, so we urge users to be aware of this if using these sources.}}. A description of the added flags can be found in Appendix \ref{app:catalogue}, where we outline the columns within the final catalogue which is released.

\section{Host Galaxy Properties}
\label{sec:hostproperties}
Having constructed our radio catalogues with matched host galaxy information we investigate the properties of the radio sources, including the luminosity distribution (Section \ref{sec:lumdist}) and the additional host galaxy properties from Stylianou et al. (in prep.). Specifically we explore the host galaxy stellar mass ($M_{*}$) and SFR distributions for our sample and make initial investigations of the redshift distribution for AGN and SFG analogues selected through whether they appear to have a radio excess in 1.4 GHz luminosity compared to expectations from the star formation rate (Section \ref{sec:m_sfr_dist}).

\begin{figure*}
    \includegraphics[width=\textwidth]{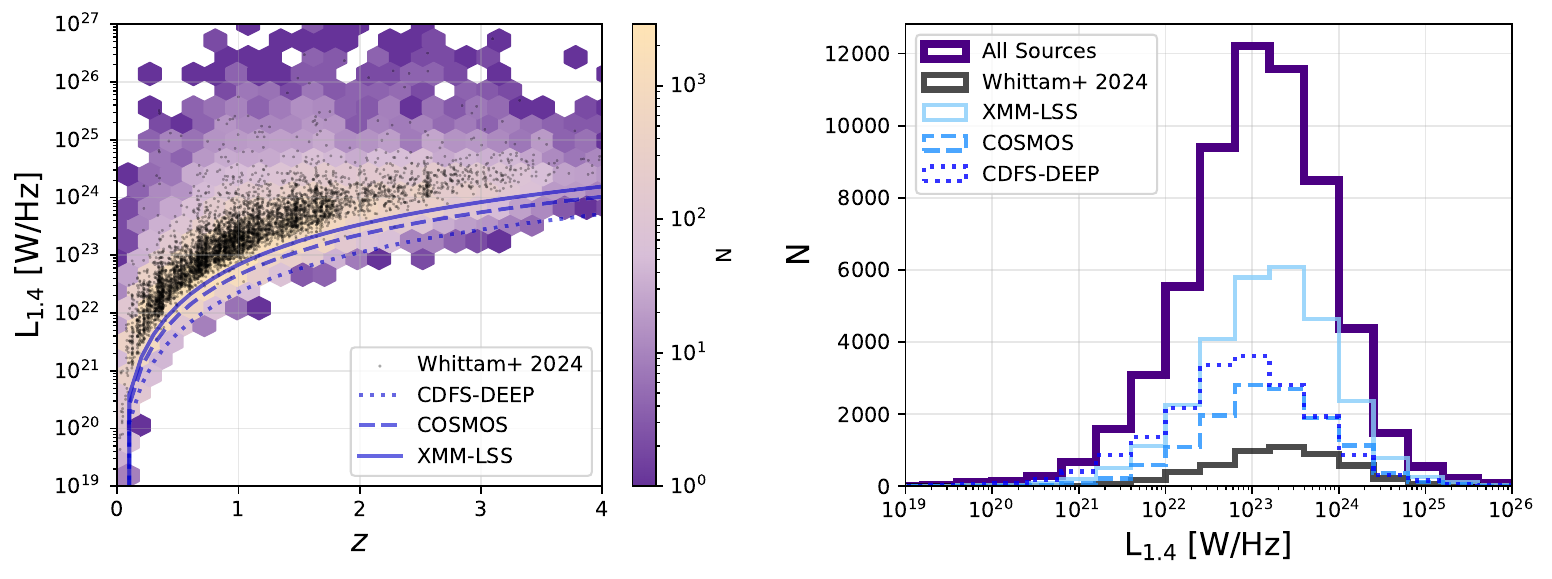}
    \caption{{The 1.4 GHz luminosity distribution of sources across the combined DR1 cross-matched catalogues. Left: the distribution of luminosity with redshift up to $z=5$, compared to the 5$\sigma$ limiting luminosity \citep[adopting the central rms values of][and an average effective frequency of 1.2 GHz]{Hale2025} for the CDFS-DEEP (black dashed), COSMOS (navy dot-dashed) and XMM-LSS (blue dotted) fields {scaled to 1.4 GHz}. Also shown (black dots) are the corresponding distributions from the MIGHTEE early science data, using the catalogue of visually cross-matched sources that were further used to classify sources \protect \citep[from][]{Whittam2022, Whittam2024}. Right: the combined distribution of luminosity for all sources with a redshift (thick purple line) and separated into those sources from the CDFS-DEEP (dark blue dotted), COSMOS (blue dashed) and XMM-LSS (light blue solid) field, as well as the MIGHTEE-ES visually inspected cross-matched catalogue (grey).}}
    \label{fig:lum_dist}
\end{figure*}

\subsection{Radio Luminosity Distributions}
\label{sec:lumdist}
First, we present a comparison of the luminosity distribution of the cross-matched sources in Figure \ref{fig:lum_dist} both as a function of redshift, as well as the full distribution of luminosities from the three fields - both with a comparison to the MIGHTEE-ES data. Combined, we have a factor of $\sim$11.6 more MIGHTEE sources with host galaxy and redshift associations than in the work of \cite{Whittam2024} ({a factor of} $\sim$2.5 more sources in the COSMOS field). This is predominantly due to the areal increase \citep[$\sim$7.5 sq deg c.f. $\sim$0.8 sq. deg in][]{Whittam2024}. {However,} there {are} also some improvements in the sensitivity both within the COSMOS field as well as from the inclusion of the CDFS-DEEP data. This is illustrated through the comparison of the sensitivity limits from the three different fields of MIGHTEE-DR1 compared to the data available from MIGHTEE-ES. For COSMOS, this improved sensitivity is as a result of the increased number of pointings in the MIGHTEE-DR1 COSMOS data (compared to the single pointing in MIGHTEE-ES) which reduces sensitivity loss in the outer edges of the field and improves image sensitivity throughout. For CDFS-DEEP, this is due to the longer integration time within a single pointing centre. 

As a result, the median luminosity being probed within each of the fields is reduced compared to the MIGHTEE-ES data. In the COSMOS and XMM-LSS fields, the median luminosities are $\sim$1.5-1.6$\times10^{23}$ WHz$^{-1}$, which is a factor of $\sim$2 lower than the median luminosity of \cite{Whittam2024} ($\sim$3$\times 10^{23}$ WHz$^{-1}$). In the deeper CDFS-DEEP field, the median luminosity {is a factor of four smaller compared to the median luminosity} MIGHTEE-ES (with a median $L_{1.4}\sim$7.4$\times 10^{22}$ WHz$^{-1}$ across the sample). This reduction in median luminosity provides a significant sample of galaxies to study the evolution of AGN and SFGs to lower luminosity sources and to higher redshifts. For example, considering the sensitivity curves in Figure \ref{fig:lum_dist}, a source with $L_{1.4}\sim 10^{23}$ WHz$^{-1}$ could be observed to $z\approx2$ in the CDFS-DEEP fields, as opposed to $z\approx1.5$ in the COSMOS field and $z\approx1$ in the MIGHTEE-ES data. This adds an extra $\sim 1-2$ Gyr of lookback time to trace the evolution of sources at this luminosity.

\begin{figure*}
    \includegraphics[width=0.48\textwidth]{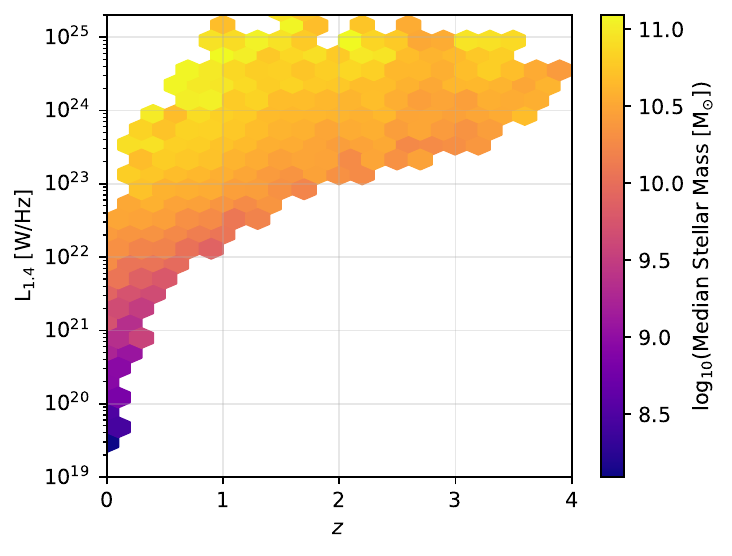} 
    \includegraphics[width=0.48\textwidth]{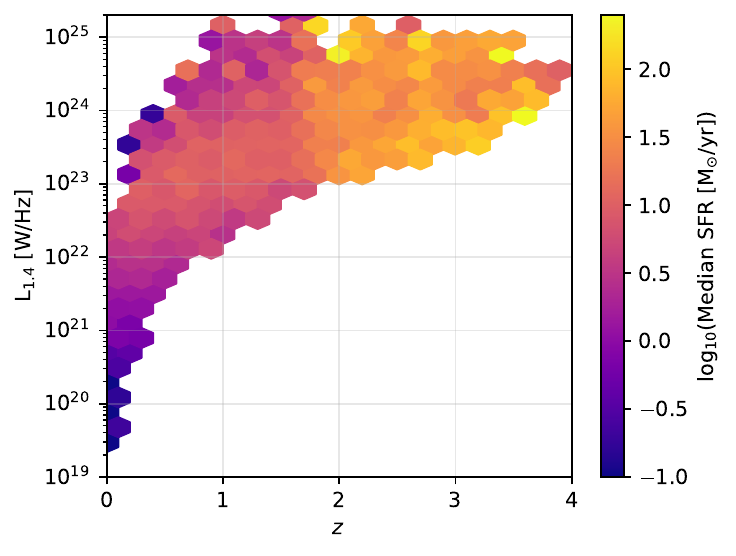}
    \caption{Median stellar mass (left) and SFR (right) for the MIGHTEE-DR1 sources with an associated host galaxy as a function of the corresponding redshift and 1.4 GHz radio luminosity. {The stellar mass and SFR are those derived from SED fitting in the work of Stylianou et al. in prep.}}
    \label{fig:mass_sfr_vs_flux_z}
\end{figure*}

\begin{figure}
    \includegraphics[width=0.5\textwidth]{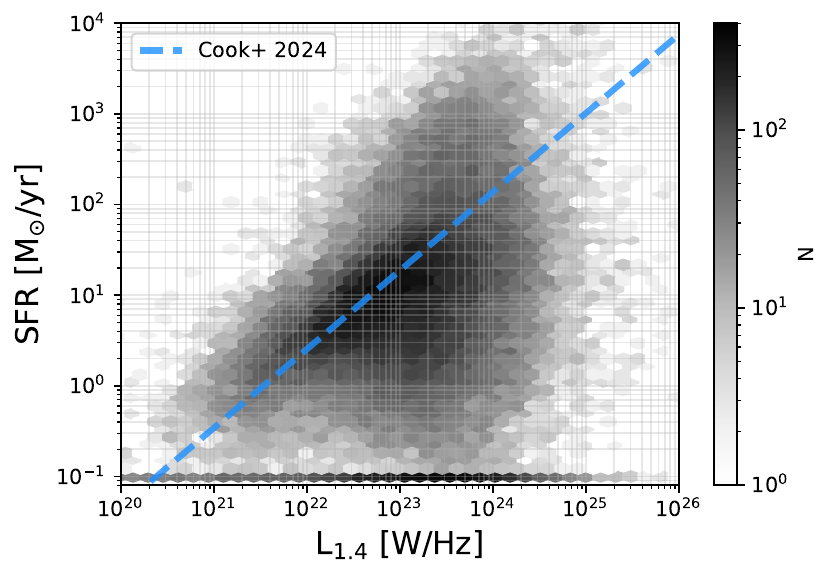} 
    \caption{Comparison of the SED measured SFR in the work of Stylianou et al. {(in prep)} as a function of the 1.4 GHz radio luminosity alongside the relation of \protect \citeauthor{Cook2024} (\citeyear{Cook2024}, blue dashed line) adopted in this work to define radio excess sources (see Section \ref{sec:m_sfr_dist}). We note that the excess of sources at SFR = 0.1 M$_{\odot}$yr$^{-1}$ (representing $\lesssim$10 per cent of the full population) represents the lower limit for the SFR in the SED fitting of Stylianou et al. {(in prep)}.}
    \label{fig:sfr_vs_L}
\end{figure}

\subsection{Redshift distributions of AGN and SFGs}
\label{sec:m_sfr_dist}
Accurate classification of radio sources into SFGs, AGN and their subsets relies upon a number of diagnostics factors from across the electromagnetic spectrum (e.g. X-ray counterparts, {very long baseline interferometry, {VLBI,}} imaging, emission lines within spectra). This includes requiring detailed modelling of the spectral energy distribution (SED) fitting to include the contribution of any AGN to the full spectrum \citep[see e.g.][]{AGNfitter, Das2024, GRAHSP}. Such classifications are essential to accurately study how SFGs and AGN evolve differently and thus the contributions of star formation and AGN activity across cosmic time. Moreover, given different physical sub-populations of AGN such as high/low excitation radio galaxies \citep[see e.g.][]{Best2012, Heckman2014, Whittam2022} and radio quiet AGN all evolve differently, it is crucial to be able to separate AGN into sub-populations and study their evolution \citep[see e.g.][]{Best2012, Pracy2016, Whittam2022, Kondapally2022, Kondapally2025}. This requires detailed modelling utilising all the information we have on these sources which is beyond the scope of this work and will be studied in future MIGHTEE {publications}.

{However, the catalogues of Stylianou et al. {(in prep)} do provide some host galaxy properties, including stellar mass and star-formation rate, were derived by Stylianou et al. {(in prep)} using the SED-fitting code \texttt{LePHARE}, with the redshifts fixed for each source\footnote{{We note that for their analysis to obtain galaxy properties, Stylianou et al. {(in prep)} fix the redshifts were fixed to those best redshift solutions using galaxy templates only (i.e. without AGN present). This is that case even if the template with AGN present provided the best fitting redshift, this is as the models with AGN can struggle to constrain the stellar mass as accurately. Therefore for some sources in CDFS ($\sim$20 per cent) where the AGN template was most appropriate and for the sources in COSMOS where the PAU redshifts were adopted, this is different to the $z$ recorded in our catalogue. However, using the same OLF criteria as in Section \ref{sec:specz}, of the total sample of sources with redshifts, sources with large OLF due to this represent $\sim$9 per cent of sources in CDFS-DEEP and <0.5 per cent of sources in COSMOS and therefore the impact of this should not be significant on our work.}}. The fits were performed using total-auto fluxes, rather than aperture fluxes, which better capture the total light of each galaxy and are therefore more suitable for estimating integrated galaxy properties. The \cite{Bruzual2003} SED templates were adopted for this step and included both solar and 20 per cent solar metallicity populations. A range of star-formation histories was considered, including a constant SFR model and exponentially declining models with timescales from 0.1 to 30 Gyr, with stellar population ages spanning 0.01-13.5 Gyr.} We do note, though, that the contamination from AGN in the SED and the lack of far-IR data in their SED fitting may bias the SFR values {and affect the ability to accurately constrain it}, this will be addressed later in the section.
We present the distribution of the median of both $M_{*}$ and SFR as a function of redshift and radio luminosity for the MIGHTEE DR1 cross-matched sources in Figure \ref{fig:mass_sfr_vs_flux_z}. As expected, the most massive galaxies appear to trace more luminous radio galaxies, which are more likely to host AGN \citep[see e.g.][]{Smolcic2017, Whittam2022, Best2023}. However, the SFR does not {exhibit smooth} trends between SFR, $z$ and luminosity. This is because {the radio sources represent} a mix of populations and whilst the SFR and radio luminosity for SFGs is correlated \citep[see e.g. relations in][]{Bell2003, Davies2017, Delhaize2017, Delvecchio2021, Cook2024}, this is not the case for AGN.  

\subsubsection{{Identifying radio excess AGN}}
In this work, we use this relationship between radio luminosity and SFR for SFGs to take a first step at studying the redshift distributions of AGN and SFGs separately in the MIGHTEE-DR1 data. For this, we utilize the SFR-L$_{1.4}$ relationship of \cite{Cook2024} which used a combination of MIGHTEE, {Deep Investigation of Neutral Gas Origins survey \citep[DINGO;][]{Rhee2023} and the Deep Extragalactic VIsible Legacy Survey \citep[DEVILS;][]{Davies2018, Davies2025}} data to measure the SFR-L$_{1.4}$ relationship to $z\sim1$ (see Figure \ref{fig:sfr_vs_L}). {We compare this to the SFRs provided by the catalogues of Stylianou et al. in prep. However,} as both the star formation history (SFH) and the choice of data used to model the SED or spectra can affect the measured SFR, we allow for variations in the relationship to the normalisation of \cite{Cook2024}. To do this, we follow the method of \cite{Mazzolari2026} and define a radio excess ratio (REX) which is calculated as the ratio of the SFR predicted by the 1.4 GHz luminosity using the relationship of \cite{Cook2024} compared to the SFR from the SED fitting. Using log$_{10}$(REX), we model the peak in this distribution ($\mu_{\textrm{log}_{10}(\textrm{REX})}$, measured to be $\sim$0.06) and use those sources with a log$_{10}$(REX) below this peak (which cannot be radio-excess sources) with a Gaussian to find a standard deviation in the relationship ($\sigma_{\textrm{log}_{10}(\textrm{REX})}$, measured to be $\sim$0.5) for SFGs around the radio excess. Those sources with:

\begin{equation}
    \textrm{REX} > 10^{\mu_{\textrm{log}_{10}(\textrm{REX})} + 2\sigma_{\textrm{log}_{10}(\textrm{REX})}}
\end{equation}

\noindent are denoted to be radio excess sources {(and are characterised here as likely radio AGN)}. Whilst previous studies have instead used 3$\sigma$ deviations in SFR-radio relations to define their radio excess criteria \citep[e.g.][]{Best2023, Mazzolari2026}, these find 3$\sigma$ deviations in the range $\sim 0.7-1$ dex, which is closer to the $\sim1.5-2 \sigma$ relations adopted in this work. However, given {that the SFRs} calculated by Stylianou et al. {(in prep)} do not include far-IR data, which can be important in constraining the SFR of galaxies, as has been used in previous studies, it is likely we might expect larger scatter in our REX. Therefore, given that 2$\sigma$ should obtain the majority ($\sim$95 per cent) of star forming sources close to the relation and that this compares {well} to previous values used in the literature, {the criteria that we use} should be sufficient for identifying star forming and radio excess sources within our sample. We note though that the SFG population will likely also include a small contribution of radio quiet AGN, whose origin of radio emission is still uncertain \citep[see e.g.][]{White2015, White2025,Njeri2026}{, jetted but non radio excess AGN \citep[see e.g.][]{Vardoulaki2021} and will also struggle to identify dusty star forming galaxies, whose emission is obscured and thus may affect their identification.}

Furthermore, as the SFR-L$_{1.4}$ relation of \cite{Cook2024} is restricted to fitting sources at $z\lesssim 1$, this may not fully account for evolution {in this relationship}, as discussed in \cite{Thykkathu2026}. Therefore, we additionally separate into radio-excess and star forming sources using an evolving REX model and fit for both $\mu_{\textrm{log}_{10}(\textrm{REX})}$ and $\sigma_{\textrm{log}_{10}(\textrm{REX})}$ within redshift bins (with $\delta z = 0.5$). This does indeed exhibit evolution in $\mu_{\textrm{log}_{10}(\textrm{REX})}$ to $\lesssim 0.9$ and, as a result of the fewer sources within increasing redshift bins there is also an increase in the scatter around the peak values. Using this evolving split will provide a check on the effect of changes to the SFR-L$_{1.4}$ models {on} our work. Numerous other relationships {between SFR and radio luminosity} have been suggested in the literature \citep[see e.g.][]{Delhaize2017, Delvecchio2021} and {we note we also investigated using these in the same way as we used the relation of \cite{Cook2024} and found that they show broadly similar redshift distributions to obtained here}.

\begin{figure*}
    \includegraphics[width=20cm, angle = 90]{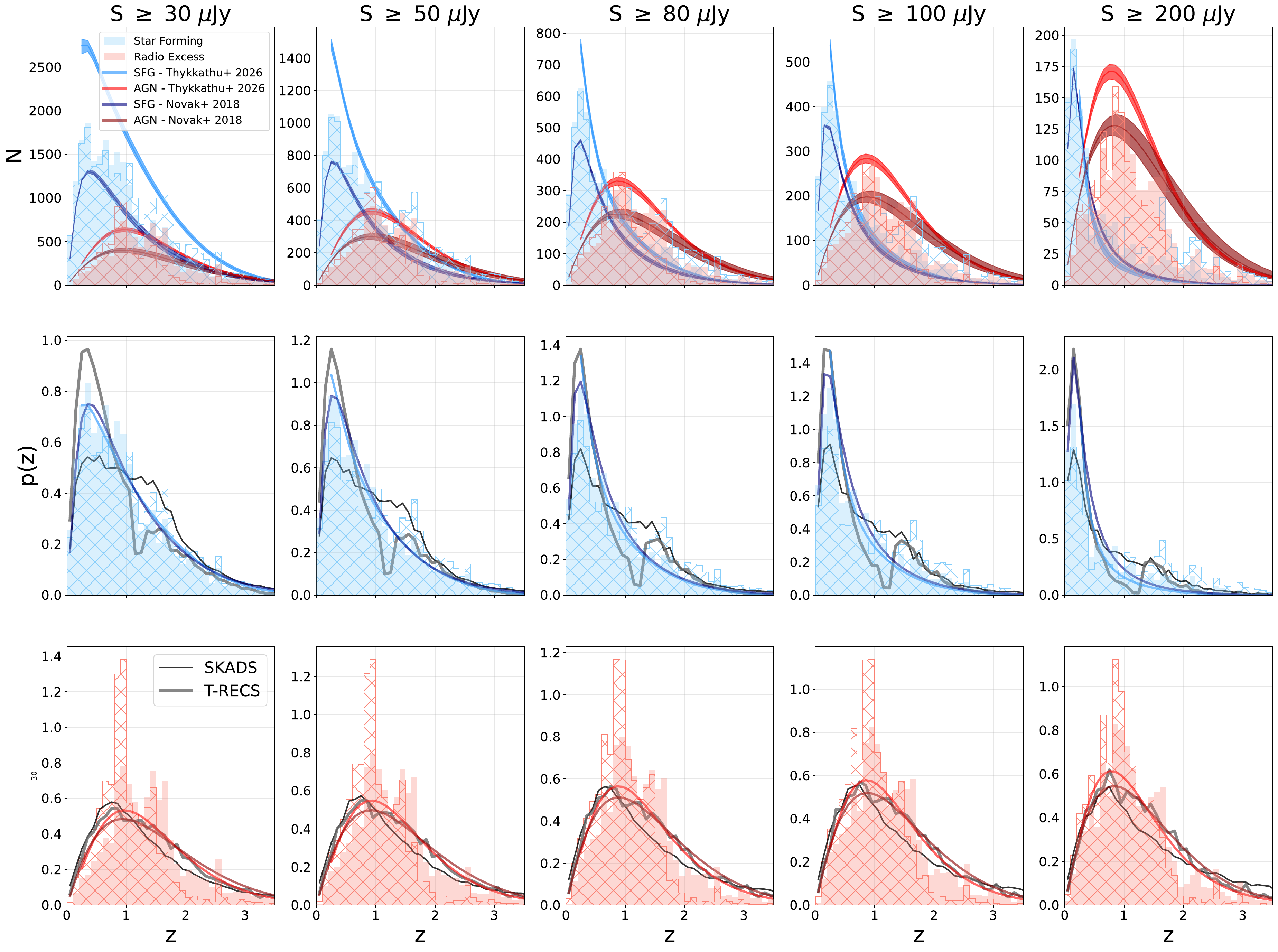}
    \caption{{Upper panel: Comparison of the observed redshift distribution of {sources identified as star-forming (blue histograms) and radio-excess (red histograms) using the selection outlined in Section} \protect \ref{sec:m_sfr_dist} for the non-evolving (filled) and evolving (hatched) REX relations. This is compared to the predicted redshift distribution from the luminosity functions of \protect \cite{Novak2018} (blue/red for SFGs and AGN) and \protect \cite{Thykkathu2026} (navy/dark red). The middle and lower panels show the normalised redshift distribution ($p(z)$) from the star forming sources and radio excess sources respectively. In the middle panel we also indicate the fraction of star forming sources from the non-evolving and evolving (indicated by (e)) models. Additionally, the two lower panels show the distributions from the simulations from SKADS \citep[][black]{Wilman2008} and T-RECS \citep[][grey]{Bonaldi2023}. The panels from left to right indicate 1.4 GHz flux limits applied of 30 $\muup$Jy, 50 $\muup$Jy, 80 $\muup$Jy, 100 $\muup$Jy and 200 $\muup$Jy respectively.}}
    \label{fig:pz_dist}
\end{figure*}

\subsubsection{{Comparison of the redshift distribution for SFG and AGN analogues}}
Adopting this split to determine radio-excess sources, we consider the redshift distributions of radio AGN and SFGs within our MIGHTEE sample and compare it to predicted distributions from luminosity functions from the VLA 3 GHz COSMOS data \citep{Novak2018} and from MIGHTEE-ES data \citep{Thykkathu2026}. We also compare to the simulated radio continuum catalogues from the SKA Design Studies \citep[SKADS;][]{Wilman2008} and from the Tiered Radio Extragalactic Continuum Simulation \citep[T-RECS;][]{Bonaldi2019, Bonaldi2023}\footnote{Using the continuum clustered output simulation of \protect \cite{Bonaldi2023}.}. We present a comparison of the expected number counts and the normalised redshift probability distributions, $p(z)$, in Figure \ref{fig:pz_dist} at a number of limiting flux densities. This allows us to compare how the redshift distributions appear over a range of flux density cuts, where we expect AGN to increase in fraction {at higher flux densities} \citep[see e.g.][]{Whittam2022, Best2023}. We present such comparisons at 30 $\muup$Jy, 50 $\muup$Jy, 80 $\muup$Jy, 100 $\muup$Jy and 200 $\muup$Jy. These are chosen as similar to the lower flux density limits which are detectable across the three fields (30 $\mu$Jy), increasing to large flux density limits which are: (i) still deep but where we believe we are $\sim$complete (50 - 80 $\muup$Jy) and SFGs dominate at the flux density limit; (ii) the flux limit where we believe the AGN and SFG are similarly contributing to the full population \citep[100 $\muup$Jy, as in][]{Whittam2022} and (iii) a higher flux density limit where AGN dominate the sample (200 $\muup$Jy). 

As shown in Figure \ref{fig:pz_dist}, whilst there are differences between adopting the non-evolving and evolving REX in classifying the MIGHTEE-DR1 sources into SFGs and radio-excess AGN, these differences are typically limited and do not substantially change the overall shape of the redshift distribution. At the larger flux densities ($S_{1.4} \geq 100 \muup$Jy)  the redshift distributions for the SFGs are in excellent agreement with both the predictions from the luminosity functions of both \cite{Novak2018} and \cite{Thykkathu2026}. At lower flux densities, {the redshift distribution of the MIGHTEE-DR1 SFG sources instead appears} to lie between the relationships of \cite{Novak2018} and \cite{Thykkathu2026}. This reflects differences in the modelling the luminosity function, where \cite{Novak2018} adopt the faint end slope from the local luminosity functions of \cite{Mauch2003}, whereas \cite{Thykkathu2026} adopt this as a free parameter. \cite{Thykkathu2026} exhibit a much larger prediction in the number of SFGs (and total sources) at the faintest flux densities. However, we note that the faint end slope in \cite{Thykkathu2026}  is driven by fitting the faintest sources in their lowest redshift bin ($z=0.2-0.4$) which they fit only to a limiting luminosity of 10$^{22}$ WHz$^{-1}$. This represents flux densities closer to $\sim 90 \muup$Jy, at the lower redshift considered ($z=0.2$), and $\sim 20 \muup$Jy at their higher redshift range ($z=0.4$). Moreover, given the MIGHTEE-ES data is noisier and at lower resolution than in this work, this could result in challenges disentangling sources at fainter flux densities or a {misestimation} of source completeness. Combined this could result in an over-prediction of the faint end slope, leading to an over-prediction of faint sources compared to that found with our MIGHTEE-DR1 catalogue. For AGN, on the other hand, both the predictions from the luminosity functions of \cite{Novak2018} and \cite{Thykkathu2026} reflect in the peak of the distribution from the MIGHTEE-DR1 data and show broad agreement in the shape of the redshift distribution, though again the work of \cite{Thykkathu2026} over-predicts the number of AGN compared to \cite{Novak2018}, with the MIGHTEE-DR1 catalogue closer to that of \cite{Novak2018}. However, given the caveats discussed above in determining the SFR of galaxies and therefore their REX, we could be mistaking radio excess AGN for SFGs, given the larger spread in our data than in previous work. Moreover, as the work of \cite{Thykkathu2026} accounts for non cross-matched sources using an assumption {that} these are all at $z>1$ (in the model compared to in this work), {it} will likely have larger numbers of sources at $z\gtrsim1$ than with our catalogue, where {do not attempt} to account for the up to $\sim$10 per cent of sources without an associated host galaxy and redshift.

Instead, focusing on the $p(z)$ distribution, we find the peak in the $p(z)$ for both the AGN and SFG analogues are in good agreement with those from the luminosity functions. This is regardless of the differences in normalisation in total {source numbers,} see Figure \ref{fig:pz_dist}. For the SFGs (star forming) populations, there is excellent agreement between the star forming galaxies in this work and those simulated SFGs in T-RECS, but {they} show more differences {in comparison to SKADS}, which appear to have a shallower peak and extend to higher redshifts. This reinforces known concerns from studies of the deep source counts from radio surveys which have highlighted the under prediction of galaxies in SKADS at the faintest flux densities, where SFGs dominate \citep[see e.g.][]{Smolcic2017, Hale2023, Hale2025}. For the AGN, these again broadly follow the models from the luminosity functions, though around $z \sim 1$ and $z \sim 1.5$ there are noticeable differences in the $p(z)$ when the evolving and non-evolving REX models are adopted, and also there {is an} excess of AGN compared to the models from the luminosity functions. This may be as a result of aliasing in the photometric redshifts, as we adopt the peak in the redshift pdf for this work, as this is the redshift adopted in Stylianou et al. {(in prep)} to obtain the SFRs. These effects might therefore be smoothed out should the full pdf be considered. The predictions of AGN distributions for both the T-RECS and SKADS simulations are both in broadly good agreement with the radio-excess in this work and both simulations peak at a similar redshift and show similar behaviour.

Overall, the redshift distributions suggest that the simple radio-excess criterion based on \cite{Cook2024} and with an evolving relation adopted in this work show broad agreement with the predicted distributions from simulations and measured luminosity functions. However, for individual sources it is imperative that more comprehensive diagnostics and SED fitting (including the far-IR) are used to accurately characterise sources. This will be addressed in future work (Jackson et al. in prep) which will provide much more comprehensive host properties for our radio sample to characterise sources and to allow for galaxy evolution studies, for different sub-populations. 

{We note that whilst this broad source classification does allow the potential to study how the AGN fraction varies as a function of host galaxy stellar mass for the near-IR sample of galaxies and to make comparisons with works such as \cite{Kondapally2025}, the simplicity in the source characterization using the REX method outlined above will lead to some misclassification of sources. Whilst misclassification is a less significant issue when the broad redshift distributions are being considered, it {can} become more important when the AGN fractions are small. This can lead to a flattening of the distribution of AGN fraction as a function of mass \citep[which is expected to be $\propto M^{2.5}$ for LERGs and  $\propto M^{1.5}$ for HERGs, see e.g.][]{Best2005, Janssen2012, Kondapally2025}, especially when the lowest mass is considered. Initial investigations by the authors identified some of these challenges and also noted differences in the relationships between the three fields studied, such as whether there was any potential redshift evolution. Such differences may arise from differences in the REX criterion applied, which uses an average relation from the combined three fields. However, given the differences in the quality of multi-wavelength data between the fields (with COSMOS having the deepest $K_s$ data) and differences in their redshift estimation, such relations may vary. As such, we leave this comparison to future work where more comprehensive source classification has been performed and where AGN {have been} sub-categorised into LERGs and HERGs, for more appropriate comparison to the work of \cite{Kondapally2025}. }

\section{Conclusions}
\label{sec:conclusions}
In this work we have presented a cross-matched catalogue for radio sources in the MIGHTEE Data Release 1 catalogues of \cite{Hale2025} over a total of 7.5 sq deg across the CDFS-DEEP (1.5 sq deg), COSMOS (1.7 sq deg) and XMM-LSS (4.3 sq deg) fields through matching to $K_s$ band catalogues from the UltraVISTA \citep{McCracken2012} and VIDEO \citep{Jarvis2013} surveys. We adopted a combined method of using statistical cross matching, through the likelihood ratio (LR) technique, and visual inspection. This combined approach minimises the need for visual inspection, allowing sources which are compact and with a high likelihood host match to be adopted, without the need for visual inspection. This is the case for $\sim80$ per cent of sources. Instead for larger sources or where the best LR counterpart is not deemed a secure enough match, these were sent for visual inspection using the Zooniverse platform and the input from members of the MIGHTEE consortium. 

Combined, this process provided host galaxy counterparts and redshifts {(within the multi-wavelength areas discussed)} for $\sim$91 per cent of sources in the CDFS-DEEP field, increasing to $\sim$92 per cent in the XMM-LSS field and $\sim$95 per cent of sources in COSMOS. Of these, close to 50 per cent of sources in the COSMOS field have spectroscopic counterparts, reducing to 30 per cent and 20 per cent of sources with spectroscopic counterparts in the XMM-LSS and CDFS-DEEP fields respectively. These fractions in CDFS-DEEP and XMM-LSS will be increased by the targetted search of spectra for MIGHTEE sources in the ORCHIDSS \citep{ORCHIDSS} survey. For those sources with counterparts, we compare the luminosity distribution to that of the MIGHTEE-ES counterpart catalogue of \cite{Whittam2024}, demonstrating that due to the areal coverage, we have an order of magnitude more sources with counterparts and for those sources with $\textrm{L}_{1.4} \sim 10^{23}$ WHz$^{-1}$ we are able to trace these over 1-2 Gyr further in lookback time due to increased in sensitivity of the data. The sky source density is additionally a factor of 2 higher compared to the LOFAR deep fields DR1 {\citep[]{Kondapally2021}, albeit with MIGHTEE-DR1 covering a smaller area and so having fewer total source numbers}. Combined, this will allow MIGHTEE-DR1 to examine the properties of {sources with} lower radio luminosities, which are crucial to understand galaxy evolution in less extreme sources. 

Finally, we use star formation rates predicted by \cite{Cook2024} based on the radio luminosity (using both an evolving and non-evolving model) compared to SFRs from SED fitting in Stylianou et al. (in prep), to define a radio-excess population (AGN analogue) and star forming sample of galaxies. Investigating the redshift distributions of these AGN and SFG analogues, we compare this to both simulated radio continuum catalogues (T-RECS and SKADS) and the inferred redshift distributions from luminosity functions from VLA 3 GHz COSMOS data \citep{Novak2018} and from MIGHTEE-ES data \citep{Thykkathu2026}. We find broad agreement in the redshift distributions with our radio excess and star forming samples, to that from the luminosity functions, though find the distribution of MIGHTEE-DR1 galaxies at the faintest flux densities are lower than predicted by MIGHTEE-ES data. As this may be a result of source confusion in the MIGHTEE-ES data and potential source blending, this emphasises that high resolution {imaging are} essential to constrain the faint end slope in the luminosity functions of SFGs {and} to accurately predict the expected distribution of the faintest sources within radio surveys. 
Whilst we have demonstrated that a radio excess cut using the host properties of the radio sources from {optical to near-IR catalogues} does allow us to broadly replicate expectations for the redshift distributions of AGN and SFGs, exact source classifications require a number of diagnostic from across the electromagnetic spectra (e.g. the identification of X-ray AGN) and using a combination of SED and spectral information. For MIGHTEE sources, improved SED modelling for AGN components and which utilise far-IR data will be essential.

\section*{Acknowledgements}
{We thank I. Smail for helpful comments on the draft manuscript.} CLH and PNB acknowledges support from the Science and Technology Facilities Council (STFC) through grant ST/Y000951/1 and from the Leverhulme Trust through an Early Career Research Fellowship. CLH, MJJ and IHW also acknowledge support from the Oxford Hintze Centre for Astrophysical Surveys which is funded through generous support from the Hintze Family Charitable Foundation.  MJJ, AAV, HP, SLJ, NS, MNT, RGV acknowledge the support from a UKRI Frontiers Research Grant [EP/X026639/1], which was selected by the ERC. SLJ also acknowledges support from the STFC consolidated grants [ST/S000488/1] and [ST/W000903/1]. MNT has been supported by funding from the European Research Council (ERC) under the European Union’s Horizon 2020 research and innovation programmes (grant agreement no. 101018897 CosmicExplorer). MJH thanks the UK STFC for support [ST/V000624/1, ST/Y001249/1]. DJBS acknowledges support from the United Kingdom’s Science and Technology Facilities Council (STFC) via grant ST/Y001028/1, and from the Leverhulme Trust via Research Project Grant RPG-2025-078. JD, LS, KKC and NN acknowledge partial research support by the National Research Foundation of South Africa (Ref Number CSUR240426216203) and an Africa-Oxford Catalyst Collaboration Grant (AfOx-290). L.V.M. acknowledges financial support from the grant CEX2021-001131-S funded by MICIU/AEI/ 10.13039/501100011033, from the grant PID2021-123930OB-C21 and PID2024-155817OB-I00 funded by MICIU/AEI/ 10.13039/501100011033 and by ERDF/EU. {LM, TFR \& MV acknowledge financial support from the Inter-University Institute for Data Intensive Astronomy (IDIA), a partnership of the University of Cape Town, the University of Pretoria and the University of the Western Cape. {LM \& MV also acknowledge support} from the South African Department of Science and Innovation’s National Research Foundation under the ISARP RADIOMAP Joint Research Scheme (DSI-NRF Grant Number 150551) and the CPRR Programme (DSI-NRF Grant Number SRUG2204254729 and SRUG22031677). The MeerKAT telescope is operated by the South African Radio Astronomy Observatory, which is a facility of the National Research Foundation, an agency of the Department of Science, Technology and Innovation. We acknowledge the use of the ilifu cloud computing facility – \url{www.ilifu.ac.za}, a partnership between the University of Cape Town, the University of the Western Cape, Stellenbosch University, Sol Plaatje University and the Cape Peninsula University of Technology. The ilifu facility is supported by contributions from the Inter-University Institute for Data Intensive Astronomy (IDIA), the Computational Biology division at UCT and the Data Intensive Research Initiative of South Africa (DIRISA). }

\section*{Data Availability}
The catalogues produced in this work will be released alongside the publication of this article.



\bibliographystyle{mnras}
\bibliography{mightee_crossmatch_dr1} 




\appendix

\section{Cross-Matched Catalogue Columns}
\label{app:catalogue}

We present the first ten lines of the catalogues released with this work in Table \ref{tab:examplecat}. Below is a descriptions of the columns listed: \\

\noindent - \texttt{\textbf{Source\_Name}} - A source name for the object given by the RA and Dec in the form JHHMMSS.SS±DDMMSS.S with a prefix of MGTDR1. The RA/Dec used for the source name is taken as the host position where available and, if unavailable, the radio position instead.

\noindent - \texttt{\textbf{RA\_radio}} - The RA of the radio centre in degrees. This is taken as the position from the \texttt{PyBDSF} catalogue, or the flux weighted centroid for objects matched within the zoo.

\noindent - \texttt{\textbf{Dec\_radio}} - The Dec of the radio centre in degrees (as above)

\noindent - \texttt{\textbf{RA\_host}} - The RA (in degrees) of the cross-matched host galaxy. If no host was assigned (either due to being in a masked region or as there was no consensus) this is denoted as -99.

\noindent - \texttt{\textbf{Dec\_host}} - The Dec (in degrees) of the cross-matched host galaxy (or -99 if no host, as above)

\noindent - \texttt{\textbf{RA\_host\_newcat}} - The RA (in degrees) of the host galaxy (if available) in the updated catalogues of Stylianou et al. {(in prep)} matched within 1 {arcsec}, see Section \ref{sec:redshifts}.

\noindent - \texttt{\textbf{Dec\_host\_newcat}} - The Dec (in degrees) of the host galaxy (if available) in the updated catalogues of Stylianou et al. {(in prep)} matched within 1 {arcsec}, see Section \ref{sec:redshifts}

\noindent - \texttt{\textbf{Flux}} - The MIGHTEE integrated flux density of the source taken 
from \texttt{PyBDSF}, or the summed flux density of the objects combined in the zoo

\noindent - \texttt{\textbf{Flux\_err}} - The error in the MIGHTEE integrated flux density of the source from \texttt{PyBDSF}, or through the propagation of uncertainties from   \texttt{PyBDSF} of the objects combined together

\noindent - \texttt{\textbf{Peak\_flux}}- The MIGHTEE peak flux density of the source taken from \texttt{PyBDSF} or the maximum peak flux density of the objects combined in the zoo

\noindent - \texttt{\textbf{Peak\_flux\_err}} - The MIGHTEE peak flux density error of the source taken from \texttt{PyBDSF} or the corresponding error associated with the maximum peak flux density of the objects combined in the zoo

\noindent - \texttt{\textbf{match\_method}} - Method used to match sources given by either: `zoo' for sources matched in the MIGHTEE zoo project; `LR' for sources matched using the Likelihood ratio method; `gaussian' for sources split into their individual Gaussian components. Sources which do not have a host are assigned a value of `NoMatch' but were all sent to the zoo for processing and those which were within masked regions are denoted as `None'. For those sources which were changed during the final checks on the catalogue (see Section \ref{sec:cat_checks}), these were flagged as `Post'.

\noindent - \texttt{\textbf{Size}} - Size of the source (in  {arcsec}) from the zoo project as discussed in \cite{Williams2019} or if the source is from the original \texttt{PyBDSF} catalogue, then the `Maj' column from \texttt{PyBDSF} is adopted

\noindent - \texttt{\textbf{Art\_prob}} - Probability of being an artefact from the MIGHTEE zoo project 

\noindent - \texttt{\textbf{Blend\_prob}} - Probability of being a blended source from the MIGHTEE zoo project 

\noindent - {\texttt{\textbf{Zoom\_prob}} - Probability of being too zoomed in from the MIGHTEE zoo project}

\noindent - {\texttt{\textbf{Hostbroken\_prob}} - Probability of having a multi-wavelength host which was split into multiple sources from the MIGHTEE zoo project}

\noindent - {\texttt{\textbf{Imagemissing\_prob}} - Probability of having a multi-wavelength image missing when looked at in the MIGHTEE zoo project}

\noindent - {\texttt{\textbf{Badclick}} - Clicks to a host which were not in the catalogue cross-matched to in MIGHTEE zoo}

\noindent - \texttt{\textbf{z\_BEST\_peak}} - Best photometric redshift from the catalogues of Stylianou+ in prep, taken as the peak in the redshift pdf ({using galaxy templates where template fitting alone})

\noindent - \texttt{\textbf{z\_phot\_hpd68\_low}} - {Lower 68th percentile around z\_BEST\_peak}

\noindent -\texttt{\textbf{z\_phot\_hpd68\_high}} - {Upper 68th percentile around z\_BEST\_peak}

\noindent - \texttt{\textbf{z\_phot\_percentile\_50}} - 50th percentile of the redshift pdf of Stylianou+ in prep

\noindent - \texttt{\textbf{z\_phot\_percentile\_16}} - 16th percentile of the redshift pdf of Stylianou+ in prep

\noindent - \texttt{\textbf{z\_phot\_percentile\_84}} - 84th percentile of the redshift pdf of Stylianou+ in prep

\noindent - \texttt{\textbf{z\_pau}} - Redshift from the PAU catalogue \cite{Alarcon2021} (COSMOS only)

\noindent - \texttt{\textbf{z\_phot}} - Best photometric redshift. For CDFS-DEEP/XMM-LSS this is the same as z\_BEST\_peak, for COSMOS it is z\_BEST\_peak else a PAU redshift was available, in which case it is z\_pau

\noindent - \texttt{\textbf{z\_spec}} - Spectroscopic redshift, if available

\noindent - \texttt{\textbf{Origin\_z\_spec}} - Flag to describe the origin of the spectroscopic redshift, where available. These redshifts were all taken from the compilation of redshifts from \cite{2022specz}\footnote{\href{https://doi.org/10.5281/zenodo.15176245} using the compilation from 20th March 2025, see also \cite{2026specz}.}. The possible labels are as follows: ACES-120604 \citep{Cooper2012}; BLAST-SPECZ \citep{Eales2009, Moncelsi2011}; {COSMOS-V1.0} \citep{Khostovan2025}; DDFDATA-V1 \citep{Zou2022}; DEIMOS-10K \citep{Hasinger2018}; DESI-DR1 \citep{DESI2025}; GAMA-DR4 \citep{Driver2022}; GAMA-G10-DR1 \citep{Davies2015}; HSU14 \citep{Hsu2014}; ICL-WHT \citep{Patel2011}; MAO12 \citep{Mao2012}; the NASA/IPAC Extragalactic Database (NED; the identifier includes a date label of YYYYMMDD for reference); OzDES-DR2 \citep{Lidman2020}; PRIMUS-DR1 \citep{Coil2011, Cool2013}; SDSS-DR16 \citep{Ahumada2020}; UDS-101018 and UDSz-140325 compilations from {Alamani} et al. of 10/10/2018 and 14/03/2025 respectively from UDS using data from \cite{Bradshaw2013, McLure2013} and \cite{Maltby2016} ; VIPERS-PDR2 \citep{Guzzo2014, Garilli2014, Scodeggio2018}; VVDS-02HR-DEEP-UDEEP \citep{LeFevre2013}; VVDS-CDFS-DEEP \citep{LeFevre2013}; hCOSMOS \citep{Damjanov2018}; zCOSMOS-BRIGHT-DR3 \citep{Lilly2007}. 

\noindent - \texttt{\textbf{z\_best\_final}} - The best redshift available for the source from either \texttt{z\_phot} or, if it is available, \texttt{z\_spec}. 

\noindent - \texttt{\textbf{Origin\_z\_best\_final}} - Flag to attribute where the redshift is taken from. This is given as one of: None; Specz; Pau\_z; Photoz

\noindent - \texttt{\textbf{EffFreq}} - Effective frequency at the source position in MHz

\noindent - \texttt{\textbf{L1400}} - 1.4 GHz scaled luminosity in WHz$^{-1}$ (assuming a spectral index, $\alpha=0.7$). Given as -99 if no redshift was available for the source

\noindent - \texttt{\textbf{Mask\_2025\_08}} - Flag to indicate whether a source was within the cross-matching area (given by a value of 1) or was within the masked region and thus not cross-matched (given by a value of 0) when the MIGHTEE zoo was started 

\noindent - \texttt{\textbf{Mask\_Matching}} - Flag to indicate whether a source was within the masked area of the catalogues of Stylianou et al. (in prep). Sources in the masked region are denoted by 0 and those in the cross-match region are given a value of 1

\noindent - \texttt{\textbf{Mask\_Final}} - Combined flag to indicate whether a source was within the area in which cross-matching took place (given by a value of 1 - taken as sources with both \texttt{Mask\_2025\_08} and \texttt{Mask\_Matching} ==1) or was within the masked region and thus not cross-matched (given by a value of 0)

\noindent - \texttt{\textbf{N\_in\_src}} - Number of initial \texttt{PyBDSF} sources which have been combined together in this final catalogue

\begin{table*}
    \centering
    \begin{tabular}{ccccccccc}
    \hline
\textbf{Source\_Name} & \textbf{RA\_radio} & \textbf{Dec\_radio} & \textbf{RA\_host} & \textbf{Dec\_host} & \textbf{RA\_host\_newcat} & \textbf{DEC\_host\_newcat} & \textbf{Flux} & \textbf{Flux\_err} \\ 
 & \textit{deg} & \textit{deg} & \textit{deg} & \textit{deg} & \textit{deg} & \textit{deg} & \textit{Jy} & \textit{Jy}  \\ \hline \hline 
MGTDR1J095852.80+022603.0 & 149.719936 & 2.434258 & 149.720009 & 2.434193 & 149.720008 & 2.434201 & 0.000020 & 0.000005 \\
MGTDR1J095852.80+023833.6 & 149.720004 & 2.642668 & -99 & -99 & -99 & -99 & 0.000034 & 0.000006 \\
MGTDR1J095852.83+021853.2 & 149.720004 & 2.314844 & 149.720151 & 2.314803 & 149.720149 & 2.314807 & 0.000034 & 0.000005 \\
MGTDR1J095852.77+021713.3 & 149.720010 & 2.286729 & 149.719886 & 2.287034 & 149.719893 & 2.287033 & 0.000017 & 0.000006 \\
MGTDR1J095852.85+021933.6 & 149.720129 & 2.326008 & 149.720212 & 2.326001 & 149.720218 & 2.325999 & 0.000038 & 0.000006 \\
MGTDR1J095852.84+012429.4 & 149.720175 & 1.408176 & -99 & -99 & -99 & -99 & 0.000184 & 0.000012 \\
MGTDR1J095852.89+023004.3 & 149.720244 & 2.500956 & 149.720405 & 2.501200 & 149.720413 & 2.501209 & 0.000024 & 0.000006 \\
MGTDR1J095852.87+015305.0 & 149.720251 & 1.884682 & 149.720329 & 1.884732 & 149.720338 & 1.884725 & 0.000043 & 0.000009 \\
MGTDR1J095852.87+025917.6 & 149.720307 & 2.988232 & -99 & -99 & -99 & -99 & 0.000044 & 0.000010 \\
MGTDR1J095852.90+024237.0 & 149.720341 & 2.710303 & 149.720436 & 2.710278 & 149.720448 & 2.710277 & 0.000054 & 0.000007 \\

    \end{tabular}
\end{table*}

\begin{table*}
    \centering
    \begin{tabular}{cccccccccc}
    \hline
\textbf{Peak\_flux} & \textbf{Peak\_flux\_err} & \textbf{match\_method} & \textbf{Size} & \textbf{Art\_prob} & \textbf{Blend\_prob} & \textbf{Zoom\_prob} & \textbf{Hostbroken} & \textbf{Imagemissing} & \textbf{Badclick} \\ 
 &  &  & &  &  &  & \textbf{\_prob} & \textbf{\_prob} &  \\ 
\textit{Jy/beam} & \textit{Jy/beam}  &  & {\textit{( {arcsec})}} &  &  &  &  &  &  \\ \hline \hline 
0.000022 & 0.000003 & LR & 5.071870 & -99 & -99 & -99 & -99 & -99 & -99 \\
0.000033 & 0.000003 & None & 5.609934 & -99 & -99 & -99 & -99 & -99 & -99 \\
0.000028 & 0.000003 & LR & 6.118757 & -99 & -99 & -99 & -99 & -99 & -99 \\
0.000014 & 0.000003 & LR & 6.132972 & -99 & -99 & -99 & -99 & -99 & -99 \\
0.000031 & 0.000003 & LR & 5.893473 & -99 & -99 & -99 & -99 & -99 & -99 \\
0.000170 & 0.000007 & None & 5.507381 & -99 & -99 & -99 & -99 & -99 & -99 \\
0.000019 & 0.000003 & LR & 6.423954 & -99 & -99 & -99 & -99 & -99 & -99 \\
0.000019 & 0.000003 & zoo & 9.601028 & 0 & 0.200000 & 0 & 0 & 0 & 0 \\
0.000043 & 0.000006 & None & 5.706223 & -99 & -99 & -99 & -99 & -99 & -99 \\
0.000050 & 0.000004 & LR & 5.559336 & -99 & -99 & -99 & -99 & -99 & -99 \\
    \end{tabular}
\end{table*}

\begin{table*}
    \centering
    \begin{tabular}{ccccccccccc}
    \hline
\textbf{Z\_BEST\_peak} & \textbf{z\_hpd68\_low} & \textbf{z\_hpd68\_high} & \textbf{z\_p50} & \textbf{z\_p16} & \textbf{z\_p84} & \textbf{z\_pau} & \textbf{z\_phot} & \textbf{z\_spec} & \textbf{Origin\_z\_spec} \\ 
 &  &  &  &  &  &  &  &  &  &  \\ \hline \hline 
0.118013 & 0.095011 & 0.139015 & 0.117116 & 0.094519 & 0.138905 & -99 & 0.118013 & 0.124988 & NED-20250320  \\
-99 & -99 & -99 & -99 & -99 & -99 & -99 & -99 & -99 & -99\\
1.093121 & 1.036115 & 1.145127 & 1.087098 & 1.021371 & 1.134961 & 1.154000 & 1.154000 & -99 & -  \\
0.654073 & 0.619069 & 0.690077 & 0.655507 & 0.620813 & 0.691617 & 0.762000 & 0.762000 & -99 & -  \\
2.294255 & 1.833204 & 2.766307 & 2.304621 & 1.842710 & 2.776072 & -99 & 2.294255 & -99 & - \\
-99 & -99 & -99 & -99 & -99 & -99 & -99 & -99 & -99 & -99\\
1.818202 & 1.754195 & 1.859207 & 1.799471 & 1.727503 & 1.845561 & -99 & 1.818202 & 0.485916 & NED-20250320  \\
1.490166 & 1.219135 & 1.782198 & 1.493301 & 1.198593 & 1.762957 & -99 & 1.490166 & -99 & -  \\
-99 & -99 & -99 & -99 & -99 & -99 & -99 & -99 & -99 & -99  \\
0.870097 & 0.836093 & 0.899100 & 0.865833 & 0.831762 & 0.895630 & 0.889000 & 0.889000 & 0.870000 & NED-20250320 \\
    \end{tabular}
\end{table*}

\begin{table*}
    \centering
    \begin{tabular}{ccccccccccc}
    \hline
\textbf{z\_best\_final} & \textbf{Origin\_z\_best\_final} & \textbf{EffFreq} & \textbf{L1400} & \textbf{Mask\_2025\_08} & \textbf{Mask\_Matching} & \textbf{Mask\_final} & \textbf{N\_in\_src} \\ 
 & &  {\textit{MHz}} & {\textit{WHz$^{-1}$}} &  &  &  &  \\ \hline \hline 
0.124988 & Specz & 1230.390000 & 7.359214e+20 &1 & 1 & 1 & 1 \\ 
-99 & None & 1210.980000 & -99 & 0 & 0 & 0 & 1 \\ 
1.154000 & Pau\_z & 1235.130000 & 1.819307e+23 &1 & 1 & 1 & 1 \\ 
0.762000 & Pau\_z & 1236.200000 & 3.602411e+22 &1 & 1 & 1 & 1 \\ 
2.294255 & Photoz & 1234.660000 & 9.909803e+23 &1 & 1 & 1 & 1 \\ 
-99 & None & 1210.690000 & -99 & 0 & 0 & 0 & 1 \\ 
0.485916 & Specz & 1225.290000 & 1.716798e+22 &1 & 1 & 1 & 1 \\ 
1.490166 & Photoz & 1234.580000 & 4.188902e+23 &1 & 1 & 1 & 1 \\ 
-99 & None & 1179.160000 & -99 & 0 & 0 & 0 & 1 \\ 
0.870000 & Specz & 1207.250000 & 1.482078e+23 &1 & 1 & 1 & 1 \\ 

    \end{tabular}
    \caption{Example 10 rows from the catalogue released alongside this work. Explanations of each column are given in Appendix \ref{app:catalogue}.}
    \label{tab:examplecat}
\end{table*}

\FloatBarrier

\noindent\textit{$^{1}$Institute for Astronomy, University of Edinburgh,  Royal Observatory Edinburgh, Blackford Hill, Edinburgh, EH9 3HJ, UK} \\
\textit{$^{2}$Astrophysics, Department of Physics, Denys Wilkinson Building, University of Oxford, Keble Road, Oxford, OX1 3RH, UK} \\
\textit{$^{3}$Department of Physics and Astronomy, University of the Western Cape, Robert Sobukwe Road, 7535 Bellville, Cape Town, South Africa} \\
\textit{$^{4}$Department of Physics, Astronomy and Mathematics, University of Hertfordshire, College Lane, Hatfield AL10 9AB, UK}\\
\textit{$^{5}$Inter-University Institute for Data Intensive Astronomy, Department of Astronomy, University of Cape Town, 7701 Rondebosch, Cape Town, South Africa}\\
\textit{$^{6}$INAF - Istituto di Radioastronomia, via Gobetti 101, 40129 Bologna, Italy}\\
\textit{$^{7}$Department of Astronomy, University of Cape Town, 7701 Rondebosch, Cape Town, South Africa}\\
\textit{$^{8}$Dipartimento di Fisica e Astronomia, Università di Bologna, Via P. Gobetti 93/2, 40129 Bologna, Italy} \\
\textit{$^{9}$UNISA Centre for Astrophysics and Space Sciences, College of Science, Engineering and Technology, University of South Africa, Cnr Christian de Wet Rd and Pioneer Avenue,Florida Park, 1709 Roodepoort, South Africa}\\
\textit{$^{10}$Green Bank Observatory, P.O. Box 2, Green Bank, WV 24944}
\textit{$^{11}$Victoria University of Wellington, Wellington, New Zealand} \\
\textit{$^{12}$South African Astronomical Observatory, PO Box  9, Observatory 7935, South Africa}\\
\textit{$^{13}$Institute for Fundamental Physics of the Universe (IFPU), Via Beirut 2, I-34151 Trieste, Italy} \\
\textit{$^{14}$SISSA - International School for Advanced Studies, Via Bonomea 265, 34136 Trieste, Italy}\\
\textit{$^{15}$INAF - Osservatorio Astronomico di Trieste, Via G. B. Tiepolo 11, I-34131 Trieste, Italy}\\
\textit{$^{16}$INFN – National Institute for Nuclear Physics, Via Valerio 2, I-34127 Trieste, Italy}\\
\textit{$^{17}$ Institute of Astronomy, University of Cambridge, Madingley Road, Cambridge CB3 0HA, UK} \\
\textit{$^{18}$Centre for Space Research, North-West University, Potchefstroom 2520, South Africa} \\
\textit{$^{19}$National Institute for Theoretical and Computational Sciences (NITheCS), Potchefstroom 2520, South Africa} \\
\textit{$^{20}$INAF, Osservatorio Astrofisico di Catania, Via Santa Sofia 78, 95123 Catania, Italy} \\
\textit{$^{21}$University College London, Department of Physics and Astronomy, Gower Street, London, WC1E 6BT, UK} \\
\textit{$^{22}$Max Planck Institute for Radio Astronomy, Auf dem H\"ugel 69, 53121 Bonn, Germany }\\
\textit{$^{23}$Jodrell Bank Centre for Astrophysics, Department of Physics and Astronomy, School of Natural Sciences, The University of Manchester, Manchester, M13 9PL, UK }\\
\textit{$^{24}$Instituto de Astrofísica de Andalucía (IAA-CSIC), Glorieta de la Astronomía s/n, 18008 Granada, Spain} \\
\textit{$^{25}${National Astronomical Observatories, Chinese Academy of Sciences, Beijing 100101, China}}\\
\textit{$^{26}${Guizhou Radio Astronomical Observatory, Guizhou University, Guiyang 550000, China}}\\
\textit{$^{27}${INAF-Osservatorio di Astrofisica e Scienza dello Spazio, Via Gobetti 93/3, 40129, Bologna, Italy}} \\
\textit{$^{28}${Department of Physics, University of Antananarivo, PO Box 906, Antananarivo 101, Madagascar}} \\
\textit{$^{29}${IAASARS, National Observatory Athens, Lofos Nymfon, 11852, Athens, Greece}} \\
\textit{$^{30}${ Thüringer Landessternwarte, Sternwarte 5, 07778 Tautenburg, Germany}} \\
\textit{$^{31}$SKA Observatory, Jodrell Bank, Lower Whitington, Macclesfield, SK11 9FT, UK}\\
\textit{$^{32}${The Observatories of the Carnegie Institution for Science, 813 Santa Barbara Street, Pasadena, CA 91101, USA}}

\bsp	
\label{lastpage}
\end{document}